\documentclass[aps,prd,reprint,nofootinbib,showkeys,showpacs,superscriptaddress,longbioliography,floatfix]{revtex4-1}

\pdfoutput=1

\usepackage{amsmath}
\usepackage{amssymb}
\usepackage[american]{babel}
\usepackage{booktabs}
\usepackage{dcolumn}  
\usepackage[T1]{fontenc}  
\usepackage{graphicx}  
\usepackage[]{hyperref}  
\usepackage[utf8]{inputenc}  
\usepackage[normalem]{ulem}
\usepackage{url}
\usepackage{multirow}
\usepackage{booktabs}
\usepackage{dashrule} 

\usepackage[normalem]{ulem}
\usepackage[dvipsnames]{xcolor}
\usepackage{float}
\DeclareUnicodeCharacter{2212}{-}

\usepackage{tikz}

\usetikzlibrary{arrows,shapes,positioning,shadows,trees}

\tikzset{
	basic/.style  = {draw, text width=2cm, drop shadow, font=\sffamily, rectangle},
	root/.style   = {basic, rounded corners=2pt, thin, align=center,
		fill=green!30},
	level 2/.style = {basic, rounded corners=6pt, thin,align=center, fill=green!60,
		text width=8em},
	level 3/.style = {basic, thin, align=left, fill=pink!60, text width=6.5em}
}

\usetikzlibrary{arrows,decorations.pathmorphing,backgrounds,fit,positioning,shapes.symbols,chains}

\newcolumntype{d}[1]{D{.}{.}{#1}}
\newcolumntype{v}[1]{D{,}{,\ }{#1}}

\makeatletter

\newcommand{\Rmnum}[1]{\expandafter\@slowromancap\romannumeral #1@}

\makeatother

\renewcommand {\arraystretch}{1.3}
\definecolor{mydarkgreen}{rgb}{0.0, 0.8, 0.0}  
\usepackage{hyperref}
\hypersetup{
    colorlinks=true, 
    linkcolor= blue, 
    urlcolor = mydarkgreen,
    citecolor=red}

\begin{document}

\title{Parametrizing the Hubble function instead of dark energy: Many possibilities}

\author{Sivasish Paul}
\email{sivasishpaul@gmail.com}
\affiliation{Department of Mathematics, Presidency University, 86/1 College Street,  Kolkata 700073, India}

\author{Raj Kumar Das}
\email{raj1996cool@gmail.com}
\affiliation{Physics and Applied Mathematics Unit, Indian Statistical Institute, 203 B. T. Road, Kolkata 700108, India}

\author{Supriya Pan}
\email{supriya.maths@presiuniv.ac.in}
\affiliation{Department of Mathematics, Presidency University, 86/1 College Street,  Kolkata 700073, India}
\affiliation{Department of Mathematics, Faculty of Applied Sciences, Durban University of Technology, Durban 4000, Republic of South Africa}

\begin{abstract}
In the present article, we propose a very simple parametrization of the Hubble function without parametrizing the dark components of the Universe. One of the novelties of the parametrization is that it may include a wide variety of the cosmological models, such as dark energy (both noninteracting and interacting fluids), modified gravity, cosmological matter creation and other known scenarios.    
The model is constrained with the latest astronomical probes from Hubble parameter measurements, three distinct versions of Type Ia Supernovae (Pantheon+, DESY5, Union3) and baryon acoustic
oscillations from Sloan Digital Sky Survey and Dark Energy Spectroscopic Instrument data
releases 1 and 2. Our results suggest a mild deviation from the standard $\Lambda$CDM cosmological model for most of the combined datasets. We also find that our model is thermodynamically consistent and performs well in the model comparison tests. 
\end{abstract}
\maketitle

\section{Introduction}\label{sec1}

The late-time accelerating expansion of the Universe is one of the striking discoveries  in cosmology and astrophysics~\cite{SupernovaCosmologyProject:1998vns,SupernovaSearchTeam:1998fmf} that significantly changed our understanding of the Universe and raised some important questions.  The reason behind this accelerating expansion is not yet clearly understood; however, there are many interesting proposals around it. The simplest formulation is to include a dark energy (DE) fluid (a hypothetical fluid with large negative pressure) within the context of Einstein's general relativity (GR),  describing the gravitational sector of the Universe. In this category, several models for DE have been proposed in the literature~\cite{Peebles:2002gy,Copeland:2006wr,Bamba:2012cp}. Alternatively, modifications of GR in various ways, widely known as modified gravity (MG) theories, can also explain this late-time accelerating expansion; see, for instance, Refs. \cite{Nojiri:2006ri,Sotiriou:2008rp,DeFelice:2010aj,Capozziello:2011et,Clifton:2011jh, Cai:2015emx, Nojiri:2017ncd,  Bahamonde:2021gfp}. Sometimes, the resulting hypothetical fluid responsible for the late-time behavior is also known as geometrical DE (GDE).  According to the observational evidence collected over the years, it has been confirmed that nearly 68\% of the total energy budget of the Universe is occupied by either DE or GDE, and the 28\% of the total energy budget of the Universe has been taken up by some nonluminous dark matter (DM) sector, which is cold or pressureless. 
Among various DE and MG candidates proposed so far, observational data suggest that a positive cosmological constant $\Lambda$ plugged within the framework of GR can successfully explain the current accelerating expansion of the Universe. Consequently, this  $\Lambda$ together with cold DM (CDM), widely known as the $\Lambda$CDM model, provides an excellent explanation to a large span of the astronomical datasets. Nevertheless, it has been observed that $\Lambda$CDM has some shortcomings, such as the cosmological constant problem~\cite{Weinberg:1988cp}, cosmic coincidence problem~\cite{Zlatev:1998tr}, and the cosmological tensions~\cite{DiValentino:2021izs,Perivolaropoulos:2021jda,Abdalla:2022yfr,Kamionkowski:2022pkx,CosmoVerse:2025txj} that have emerged recently. This suggests that an alternative cosmological scenario beyond $\Lambda$CDM, irrespective of whether this is constructed within GR or beyond GR, is welcome.

In the present article, we consider a novel parametrization of the Hubble function of the Friedmann-Lema\^{i}tre-Robertson-Walker (FLRW) universe, which describes a possible revision of the $\Lambda$CDM cosmology. In $\Lambda$CDM cosmology, the Hubble function evolves with the cosmological redshift $z$ as $H^2/H_0^2 = \Omega_{m0}(1+z)^3 + (1-\Omega_{m0})$,  where $\Omega_{m0}$ is the present-day value of the matter density parameter, and $H_0$ is the Hubble constant. In the
$w$CDM cosmology, which is the simplest generalization of the $\Lambda$CDM cosmology introducing a constant equation of state (EoS) of DE, $w ~(\neq -1)$, the Hubble function becomes $H^2/H_0^2 = \Omega_{m0}(1+z)^3 + (1-\Omega_{m0}) (1+z)^{3 (1+w)}$. 
As $\Lambda$CDM and $w$CDM  models represent simple cosmological structures (the dark fluids are separately conserved in both the models), it is expected that they will lead to simple Hubble functions. 
However, this is not true, in general, because there are several cosmological scenarios that lead to simple expansion laws of the Hubble function. For instance, in the context of an interacting scenario between CDM and DE, if the interaction rate is proportional to the energy density of the DE,  then, as explicitly shown in ~\cite{Benisty:2024lmj}, the Hubble function may take the form ${H^2}/{H_0^2} = \Omega_0 (1+z)^{3} + (1 - \Omega_0) (1+z)^{\alpha}$, where $\Omega_0$ and 
$\alpha$ are constants. It is important to note that $\Omega_0$ may not correspond to the present-day value of the density parameter of one of the dark fluids, and similarly $\alpha$ does not necessarily correspond to either the DE EoS or the strength of the interaction.  
On the other hand, if CDM interacts with the vacuum energy density and the interaction rate is proportional to the energy density of CDM, then the Hubble expansion law becomes ~\cite{Pan:2025qwy}  ${H^2}/{H_0^2} = \Omega_0 (1+z)^{3+\delta} + (1 - \Omega_0)$, where again $\Omega_0$ and $\delta$ are just constants. 
On the other hand, in the context of some specific MG models, the Hubble function may evolve in a similar
fashion~\cite{Nojiri:2007cq,Nojiri:2009kx, Dunsby:2010wg,Chakraborty:2025qlv}. Moreover, similar evolution of the Hubble function has been noticed in matter creation models~\cite{Alcaniz:1999hu,Lima:2009ic,Lima:2015xpa,Nunes:2016aup}. Therefore, a direct parametrization
of the Hubble function can be regarded as a novel one since it may correspond to a wide variety of the cosmological scenarios.

Motivated by this, in this work, we propose a new parametrization of the Hubble function as follows: 
${H^2}/{H_0^2} = \Omega_0 (1+z)^{3+\delta} + (1 - \Omega_0) (1+z)^{\alpha}$ where $\delta$, $\alpha$ and $\Omega_0$ are simply constants.  We constrain this generalized parametric scenario using the recent cosmological probes, including the measurements of the Hubble parameter at different redshifts~\cite{Yang:2023qsz}, baryonic acoustic oscillations (BAO) data from two different astronomical missions, namely, the  Dark Energy Spectroscopic Instrument (DESI)~\cite{DESI:2024mwx,DESI:2025zgx} and Sloan Digital Sky Survey (SDSS)~\cite{eBOSS:2020yzd}, as well as three different compilations of the Type Ia Supernovae (PantheonPlus~\cite{Scolnic:2021amr, Brout:2022vxf}, Union3~\cite{Rubin:2023ovl}, and DESY5~\cite{DES:2024jxu}). 
Additionally, we explore the laws of thermodynamics for this model, aiming to understand whether the model is consistent with the thermodynamical interpretations at the cosmological level~\cite{Cai:2005ra,Akbar:2006er,Akbar:2006kj,Jamil:2009eb, Bandyopadhyay:2010zz,Karami:2010zz,deHaro:2015hdp,Mahith:2018uab, Saridakis:2020cqq, Nojiri:2022nmu, Odintsov:2024hzu}. Finally, we perform an assessment of the model in terms of two statistical measures, namely the information criteria and the Bayesian evidence analysis. These two measures offer a quantitative view on the possible improvement of the proposed model over the $\Lambda$CDM one.

The article is organized as follows.  In Sec.~\ref{sec-2}, we introduce the parametric model at the background level, while in Sec.~\ref{subsec-perturbed}, we present the behavior of the model in the linear perturbation regime.  Section~\ref{data} presents the observational datasets and the methodology of our analysis. In Sec.~\ref{sec-results}, we present the results extracted from this scenario, in particular, the observational constraints (Sec.~\ref{sec-cons}), thermodynamical interpretation (Sec.~\ref{sec-thermo}), and the model comparison statistics using the information criteria and the Bayesian evidence analyses (Sec.~\ref{sec-model-comparison}). Finally, in Sec.~\ref{sec-summary}, we summarize our work.

\section{Parametrization of the Hubble function}
\label{sec-2}

We consider the spatially flat FLRW line element for our Universe, described by the line element $ds^2 = -dt^2 + a^2 (t) (dx_{1}^2 + dx_{2}^2 +dx_{3}^2)$, where $a(t)$ refers to the expansion scale factor of the Universe in terms of the cosmic time $t$; and ($t, x_1, x_2, x_3$) are the comoving coordinates. Assuming GR as the gravitational sector of the Universe, the Hubble expansion $H = (1/a) \times da/dt$  of the $\Lambda$CDM model  follows the simplest evolution,

\begin{eqnarray}\label{LCDM-Hubble}
    H^2/H_0^2 =  \Omega_{m0} \left(1+z\right)^{3} + (1- \Omega_{m0}),
\end{eqnarray}
where $H_0$ is the current value of the Hubble parameter $H$, and $\Omega_{m0}$ is the matter density parameter  at the present  epoch. For the analysis, we express redshift $z$ in terms of the scale factor $a$ as $1+z = 1/a$, where we set $a = 1$ at the present epoch.

The next generalization of the $\Lambda$CDM model is the $w$CDM cosmology in which DE has a constant EoS, $w \neq -1$, and both DE and CDM are separately conserved. In this case, the Hubble parameter evolves as follows:

\begin{eqnarray}\label{wCDM-Hubble}
    H^2/H_0^2 =  \Omega_{m0} \left(1+z\right)^{3} + (1- \Omega_{m0}) \left(1+z\right)^{3 (1+w)}. 
\end{eqnarray}
In both $\Lambda$CDM and $w$CDM cosmologies, it is assumed that CDM and DE are separately conserved, and the resulting Hubble equation assumes a simple form, as shown in Eq. \eqref{LCDM-Hubble} for $\Lambda$CDM and Eq. \eqref{wCDM-Hubble} for $w$CDM. However, it is important to remark that 
a Hubble equation resembling the form of Eq.~\eqref{LCDM-Hubble}  or Eq.~\eqref{wCDM-Hubble} can emerge from entirely different theoretical frameworks. For instance, a cosmological scenario allowing an interaction between DE and CDM  may lead to a resulting Hubble function that closely mimics Eq.~\eqref{wCDM-Hubble}~\cite{Benisty:2024lmj,Pan:2012ki}.  By ``closely mimics'', we mean that the Hubble equation may take the form $H^2/H_0^2 = M (1+z)^{3} + (1-M) (1+z)^y$, where $y$ and $M$ are constants, and $M$ may not correspond to the density parameter of one of the participating fluids at the present epoch [see, for instance, Eq. (6) of Ref.~\cite{Benisty:2024lmj} excluding baryonic contributions, Eq. (35) of Ref.~\cite{Pan:2012ki}, or Eq. (19) of Ref.~\cite{Pan:2016ngu}].\footnote{Similar evolution has been observed in various modified gravity models~\cite{Nojiri:2009kx,Dunsby:2010wg,Chakraborty:2025qlv}. This indicates that even if the expansion history of the Universe follows a $\Lambda$CDM or $w$CDM-like evolution, it may not imply that the gravitational sector of the Universe corresponds to GR.} This, together with the work of \cite{He:2024jku}, motivates us to propose a generalized parametrization of the Hubble function of the form~\footnote{We will refer to this as our model throughout the article.}
\begin{eqnarray}\label{model-new}
\frac{H^2}{H_0^2} = \Omega_{0} \left(1+z\right)^{3+\delta} + (1- \Omega_{0}) \left(1+z\right)^{\alpha},
\end{eqnarray}
where  $\delta$, $\alpha$, $\Omega_0$ are constants.\footnote{In the most general framework, $\delta$ and $\alpha$ can be treated as arbitrary time-dependent quantities. However, this freedom leads to degeneracy in their possible forms, allowing multiple valid choices. As a result, adopting time-dependent parametrizations for $\delta$ and $\alpha$ demands a careful and systematic analysis, which is essential for a deeper understanding of the cosmic evolution governed by such a Hubble expansion. } For $(\delta, \alpha)= (0, 0)$, the model described by Eq. \eqref{model-new} mimics the $\Lambda$CDM-like evolution. For $\delta =0$, our model mimics the $w$CDM-like evolution. Although our model (\ref{model-new}) looks very similar to that of Ref. \cite{He:2024jku} except for the presence of $\delta$, it is actually not. This is because the parameter $\Omega_{0}$ in our model may not correspond to the density parameter of the standard matter sector at present; it may involve the density parameters of the participating fluids at present, together with other model parameters. This can be better understood from Eq. (6) of \cite{Benisty:2024lmj} (also see Refs.~\cite{Pan:2012ki,Pan:2016ngu,Sharov:2017iue}) where $\Omega_0$ is shown to depend on the DE EoS and the coupling parameter(s) of the interaction. As we shall elaborate later, our model exhibits several notable features that, to the best of our knowledge, have not been reported previously.  

On the other hand, our model can also be derived from a quintessence scalar field Lagrangian $\mathcal{L}(\phi,\partial_\mu \phi) = - \frac{1}{2} g^{\mu\nu}\partial_\mu \phi \partial_\nu \phi - V(\phi)$,  with the following potential~\cite{Rubano:2001xi}:
\begin{align}
    V(\phi) \propto \left( \sinh\left( \frac{(3+\delta - \alpha)}{M_{\text{pl}}\sqrt{\alpha}} \left( \phi - \phi_0 \right) \right) \right)^{-2\alpha/(3+\delta - \alpha)}, 
\end{align}
 where $\phi$ is the scalar field, $\phi_0$ is a constant and $\delta$, $\alpha$ are the model parameters of the Hubble function ($M_{\text{pl}} = \frac{1}{\sqrt{8\pi G}}$ is the reduced Planck mass).

Now we shall describe the main novelties of the model. The parametrized form of the Hubble function in Eq. \eqref{model-new} offers  several new insights. If we interpret Eq. \eqref{model-new} as  a result of two noninteracting fluids, where the first term on the right-hand side of Eq. \eqref{model-new} corresponds to the evolution of a DM-like fluid, and the second term on the right-hand side of Eq. \eqref{model-new} corresponds to the DE evolution, then one may argue that the DM is not purely cold; rather it has a noncold nature with EoS $\delta/3$,  while the DE EoS is $w = -1 + \alpha/3$. In this scenario, $\Omega_0$ becomes identical to $\Omega_{m0}$, and now the model approaches that of Ref. \cite{He:2024jku}; 
however, it is not exactly the same, as our model includes an additional free parameter, $\delta$, which characterizes the noncold nature of DM.\footnote{The possibility of noncold nature of DM is getting significant attention in recent times~\cite{Yao:2025kuz,Braglia:2025gdo,Li:2025dwz,Wang:2025hlh,Li:2025eqh,Chen:2025wwn,Abedin:2025dis,Kumar:2025etf,Yang:2025ume,Yao:2023ybs}. } 
On the other hand,  as explicitly described in Refs. \cite{Pan:2012ki,Pan:2016ngu,Sharov:2017iue,Benisty:2024lmj}, 
the solution of the Hubble function as given in Eq. \eqref{model-new}  may also result from an interacting two-fluid system, where the exponents of $(1+z)$ might be related to the coupling parameter(s) and/or EoS of the DE, depending on the coupling function  that leads to Eq.~\eqref{model-new}. Furthermore, this model gives rise to several additional and noteworthy implications. To illustrate these, we rewrite Eq. \eqref{model-new} by introducing an analytic function $g(z)$ as follows\footnote{In this article, we use $g(z)$ to denote a function of redshift $z$ and $g(a)$ to denote the same function in terms of the scale factor $a$. These notations are used synonymously.}: 

\begin{align}\label{model-new-alternative}
\frac{H^2}{H_0^2} = \Omega_{0} \left(1+z\right)^{3+\delta} + g(z) +  (1- \Omega_{0}) \left(1+z\right)^{\alpha} - g(z),
\end{align}
where the choice of $g(z)$ is not unique, and for simplicity, we assume that the above evolution corresponds to a two-fluid system, as follows\footnote{Note that Eq. \eqref{model-new} results from a system involving multiple fluids; however, just to make the underlying scenario simple, without any loss of generality, we consider only a two-fluid system in this article. }: 

\begin{eqnarray}\label{eqn-rho1-rho2}
\left\{\begin{array}{ccc}
\widetilde{\Omega}_{1} \equiv \rho_1/\rho_{c0} &=& \Omega_{0} \left(1+z\right)^{3+\delta} + g(z),\\
\widetilde{\Omega}_2  \equiv \rho_2/\rho_{c0} &=& (1- \Omega_{0}) \left(1+z\right)^{\alpha} - g(z), \end{array}\right.  
\end{eqnarray}
where $\rho_1$, $\rho_2$ are the energy densities of the fluids, with $p_1$, $p_2$ being their respective pressure components ($\rho_{c0} = 3H_0^2/8\pi G$ is the critical energy density at the present time). 
The conservation equation of the total system is 

\begin{eqnarray}\label{cons-total}
    \dot{\rho}_1 + \dot{\rho}_2 + 3 H (1+w_1) \rho_1 + 3 H (1+w_2) \rho_2 =0,
\end{eqnarray}
where $w_1 = p_1/\rho_1$ and $w_2 = p_2/\rho_2$ are the equation of state parameters of the fluids, and they could be either constant or dynamical. 
Now, if the fluids do not interact with each other, then the conservation equation of the individual fluid becomes $\dot{\rho}_1 + 3 H (1+w_1) \rho_1 =0$ and $\dot{\rho}_2 + 3 H(1+w_2) =0$, and consequently, one finds that, $w_1 = -1 - (a/3\rho_1) \times d\rho_1/da$ and $w_2 = -1 - (a/3\rho_2) \times d\rho_2/da$. Now, using the expressions of the energy densities as given in Eq. \eqref{eqn-rho1-rho2}, one arrives at

\begin{align}
    w_1 &= -1 + \frac{1}{3} \times \frac{(3+\delta) \Omega_0 a^{-(3+\delta)} - a~dg(a)/da}{\Omega_{0} a^{-(3+\delta)} + g(a)},\\
    w_2 &= -1 +  \frac{1}{3} \times \frac{\alpha (1-\Omega_0) a^{-\alpha} + a ~ dg(a)/da}{(1- \Omega_{0}) a^{-\alpha} - g(a)}.
\end{align}
Therefore, for a specific $g(a)$, one can explicitly determine the EoS parameters. 
On the other hand, if  Eq. \eqref{model-new} [alternatively, Eq. \eqref{model-new-alternative}] describes a solution of the interaction between the fluids, then the conservation equation of the  fluids can be written as 

\begin{eqnarray}
\frac{d\rho_1}{dt} + 3 H (1+w_1) \rho_1 &=& - Q,\label{cons-1}\\
\frac{d\rho_2}{dt} + 3 H (1+w_2) \rho_2 &=& Q,\label{cons-2}
\end{eqnarray}
where $Q$ is the interaction function between these fluids. For $Q > 0$, energy transfer occurs from the first fluid [characterized by ($\rho_1$, $p_1$)] to the second fluid [characterized by ($\rho_2$, $p_2$)], while for $Q <0$, the direction of energy transfer is reversed. 
Now, plugging the expressions of $\rho_1$ and $\rho_2$ from Eq. \eqref{eqn-rho1-rho2} into  Eq. \eqref{cons-total}, one arrives at

\begin{align}
g (a)  &= \left(\frac{\delta/3 -w_1}{w_1 - w_2}\right)\Omega_{0}  a^{-(3+\delta)} -\nonumber\\
&\hspace{1.0cm} \left(\frac{1+w_2 - \alpha/3}{w_1 - w_2} \right) (1-\Omega_{0}) a^{-\alpha}.
\label{eqn-fa}
\end{align}
With this expression for $g$, it is now clear that the evolution of the energy densities, as given in Eq. \eqref{eqn-rho1-rho2}, is different from their independent evolution without any interaction, although the resulting picture follows Eq. \eqref{model-new}. The above expression indicates that without specifying an interaction function $Q$, one can derive the evolution equations for the participating fluids. Also, we can provide an interpretation of the interaction function $Q$ based on the expression of $g(a)$ in Eq. \eqref{eqn-fa}. We can write $g(a)$ as
\begin{eqnarray}
g (a)  = A\Omega_{0} a^{-(3+\delta)} -  B(1-\Omega_{0}) a^{-\alpha},
\label{eqn-fa1}
\end{eqnarray}
where $A =\left(\frac{ \delta/3-w_1}{w_1 - w_2}\right) $ and $B = \left(\frac{1+w_2 - \alpha/3}{w_1 - w_2} \right)$. Now, subtracting Eq. \eqref{cons-1} from Eq. \eqref{cons-2} and using the expressions for $\rho_1$ and $\rho_2$ as given in Eq. \eqref{eqn-rho1-rho2},  one can derive a general form of the interaction function as follows:

\begin{align}
    Q &= \frac{H}{2}\bigg(\frac{3H_0^2}{8\pi G}\bigg)\bigg[\bigg(\delta-3w_1\bigg)\Omega_{0} a^{-(3+\delta)}  \nonumber \\ 
    &\hspace{1.1cm} +\bigg(3+3w_2-\alpha\bigg)(1-\Omega_{0}) a^{-\alpha}  \nonumber \\
    &\hspace{1.5cm} -\bigg(2a\frac{dg(a)}{da} + 3(w_1+w_2+2)g(a)\bigg)  \bigg],
\end{align}
which assumes the following general form of the interaction:

\begin{eqnarray}\label{eqn:inter_func_model}
    Q = \frac{H}{2} \Bigg(\xi (a)~\rho_1+\eta (a)~\rho_2\bigg),
\end{eqnarray}
where $\xi (a) = \xi_1 + (2a/(w_1 -w_2)) \times (dw_1/da)$ and  $\eta (a) = \xi_2 + (2a/(w_1 -w_2)) \times (dw_2/da)$ are the coupling parameters of the interaction function  in which $\xi_1$ and $\xi_2$ are constants taking the forms
\begin{align}
   \xi_1 &=-3w_1+ \frac{\delta(1+B)(1+2A)+(3-\alpha)(A(1+2B))}{1+A+B},\\
\xi_2 &=3w_2+ \frac{\delta B(1+2A)+(3-\alpha)((1+A)(1+2B))}{1+A+B}. 
\end{align}
Thus, Eq.~\eqref{eqn:inter_func_model} represents a very general interaction function whose coupling parameters could, in principle, be dynamical with the cosmic evolution. Under certain conditions, the coupling parameters could be constant.  Now, notice that depending on the model parameters, Eq.~\eqref{eqn:inter_func_model} could lead to a wide variety of global interaction functions, such as $Q \propto H \rho_1$,  $Q \propto H \rho_2$, or $Q \propto H (\rho_1 +\rho_2)$, where the coupling parameters could be either constant or dynamical. Looking at the existing literature on interacting cosmological models, if $\rho_1$ corresponds to the energy density of CDM and $\rho_2$ corresponds to the energy density of DE, then one arrives at some well-known global interaction functions \cite{Farrar:2003uw, Valiviita:2009nu, Li:2011ga, Clemson:2011an, Wang:2016lxa, DiValentino:2017iww, Yang:2018euj, Yang:2018uae, Lucca:2021eqy, Pan:2023mie, Giare:2024smz, Silva:2025hxw,Shah:2025ayl,Pan:2025qwy}. See Refs.~\cite{Wang:2016lxa,Yang:2018euj, Yang:2018uae, Pan:2023mie, Giare:2024smz, Silva:2025hxw,Shah:2025ayl,Pan:2025qwy}, in which these models have been studied assuming the constant coupling parameter, and see Refs. \cite{Li:2011ga,Wang:2018azy,Yang:2019uzo,Yang:2020tax}, where the variable coupling parameter has been assumed.  
Now, concerning the dynamical nature of the EoS parameters, there is, in principle, no such guiding theory that can determine their actual forms. Based on the time-dependent nature of $w_1$ and $w_2$, the dynamical nature of the coupling parameters, namely, $\xi (a)$ and $\eta (a)$ will be determined. Although many phenomenological forms of the EoS parameters can be considered, in the rest of the article we shall perform the analysis assuming a constant EoS of the fluids, which results in constant coupling parameters for the interaction functions.

In addition, with the parametrization of the Hubble parameter, the total EoS can be derived using the Friedmann equations. For the spatially flat FLRW universe, the  Friedmann equations are given by $3H^2 = \kappa^2 \rho_{\rm total}$ and $2~(dH/dt) + 3H^2 = - \kappa^2 p_{\rm total}$ ($\kappa^2 = 8 \pi G$ is Einstein's gravitational constant, in which $G$ is Newton's gravitational constant), where $\rho_{\rm total}$ and $p_{\rm total}$ are, respectively, the total energy density and total pressure of the energy content of the Universe.  From these equations, the total EoS parameter, defined as $w_{\rm t} = p_{\rm total}/\rho_{\rm total} = -(1 + 2 (dH/dt)/3H^2 )$, can be explicitly written in terms of redshift $z$ as

\begin{align}
w_{t} &=  -1 +\frac{1}{3}\Bigg(\frac{(3+\delta)\Omega_0(1+z)^{(3+\delta)}}{\Omega_0(1+z)^{(3+\delta)} + (1-\Omega_0)(1+z)^{\alpha}} + \nonumber \\
&\hspace{1.5cm}\frac{\alpha(1-\Omega_0)(1+z)^{\alpha}}{\Omega_0(1+z)^{(3+\delta)} + (1-\Omega_0)(1+z)^{\alpha}} \Bigg ), 
\label{eq-eos}
\end{align}
which allows for the estimation of the Universe's  effective nature at the present epoch. The current value of $w_{\rm t}$ is given by $w_{\rm t,0} = -1 + (1/3) \times \bigl[ (3+\delta) \Omega_{0} + \alpha (1-\Omega_0) \bigr]$. This indicates that depending on the model parameters, $w_{\rm t,0}$ could lie in the quintessence regime ($w_{\rm t,0} > -1$), in the phantom regime ($w_{\rm t,0} < -1$), or correspond to  $w_{\rm t,0} = -1$. 

\begin{figure*}
    \centering
    \includegraphics[width=1.03\textwidth]{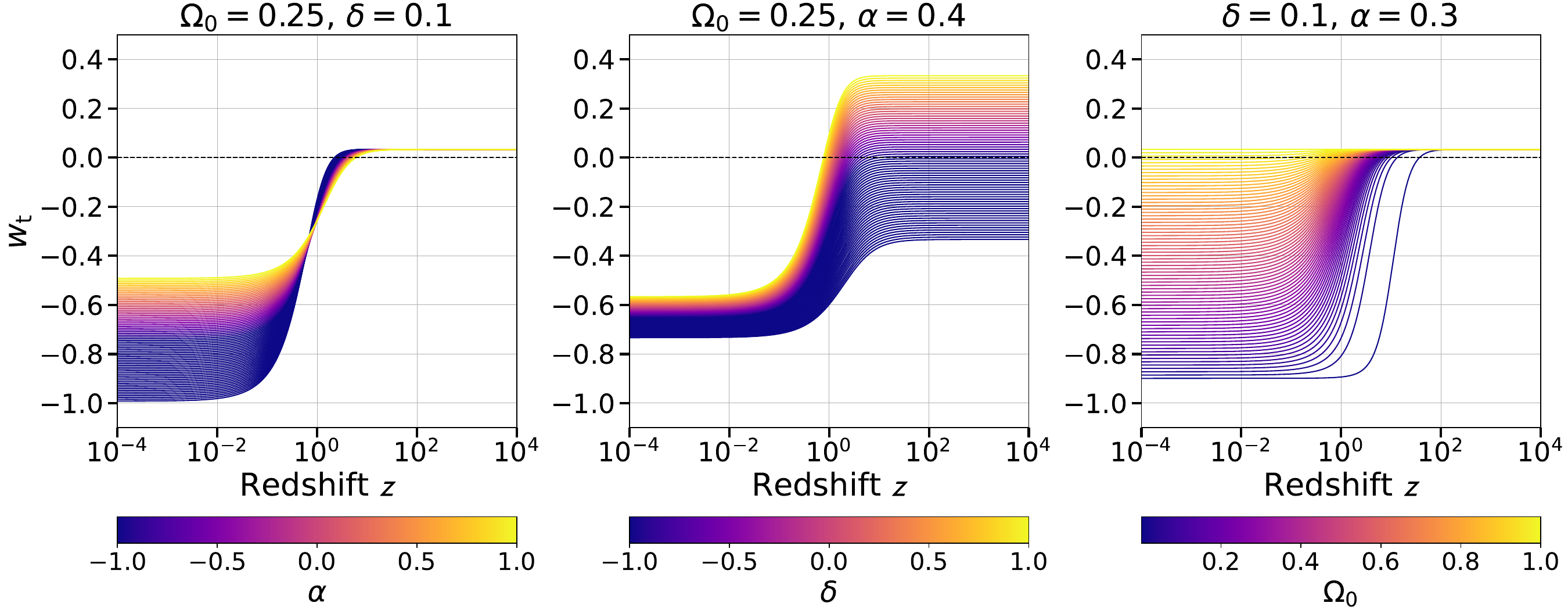}
    \caption{Redshift evolution of the total equation of state parameter $w_t$ for different values of the model parameters \{$\alpha, \delta, \Omega_0 $\}. Despite variations in the parameter choices, a quintessential behavior is consistently observed at the present epoch.}
    \label{fig:w_tot_var}
\end{figure*}

To further investigate the effective nature of the universe, the redshift evolution of the total EoS parameter $w_{t}(z)$ is analyzed by varying each parameter $\alpha$, $\delta$, and $\Omega_0$ individually, while keeping the other two fixed, as depicted in Fig.~\ref{fig:w_tot_var}. It is observed that the variation of parameters $\alpha$ and $\Omega_0$ leaves a significant imprint on the present-day value of $w_{t}$, with minimal influence on early times. In contrast, the variation of the $\delta$ parameter impacts the behavior of $w_t$ across both early and late epochs. While maintaining a constant positive $w_t$ at early times, a shift toward the phantom barrier can be observed at late epochs with decreasing values of $\alpha$ and $\Omega_0$. A similar trend is also observed with the variation of $\delta$ at the current epoch, with the additional feature of a shift from positive to negative at early times. However, when $\Omega_0 \to 1$, we observe a hint of positive $w_t$ at the present epoch. Despite the different choices of $\alpha$, $\delta$, and $\Omega_0$, the specific parametrization of the Hubble parameter, as described in Eq. (\ref{model-new}), suggests a quintessence-like behavior of the effective EoS at the current time. A detailed overview of the constrained value of $w_t$ at the present time, denoted as $w_{t,0}$, using the recent observational data, is provided in Sec.~\ref{sec-cons}. 

\subsection{Analysis at the Linear perturbation regime}
\label{subsec-perturbed}

In this section, we analyze the growth of cosmic structures both in the absence and presence of interactions between the two considered fluids, characterized by their energy densities, namely,  $\rho_1$ and $\rho_2$. We assume  that the fluid $\rho_1$ is evolving as DM, while the fluid $\rho_2$ is evolving as DE. We focus on the case of isentropic scalar perturbations, adopting the conformal Newtonian gauge, where the perturbed metric is expressed as \cite{Ma:1995ey}:
\begin{eqnarray}
    ds^2 = a^2(\tau)(-(1 + 2\psi)d\tau^2 + (1 - 2\phi)\delta_{ij}dx^i dx^j),
\end{eqnarray}
where $a(\tau)$ is the scale factor as a function of conformal time $\tau$ and $\phi=\phi(x^\mu)$ and $\psi = \psi(x^\mu)$ are the scalar potentials representing the gravitational potential and spatial curvature, respectively.

\subsubsection{Noninteractive Scenario}\label{subsec:non-int_per}

In the noninteracting scenario, where there is no energy transfer between the two fluids such that $g(a) = 0$, the evolution equations in Fourier space $k$ can be expressed in the conformal Newtonian gauge by incorporating the perturbed metric and perturbed energy-momentum tensors as follows \cite{Ma:1995ey}:

\begin{align}
    \dot{\delta_i} &= - (1+ w_i)(\theta_i - 3 \dot{\phi}) - 3 \frac{\dot{a}}{a} \left(\frac{\delta P_i}{\delta \rho_i} - w_i\right) \delta_i, \label{eqn:den_per}\\
    \dot{\theta_i} &= - \left(\frac{\dot{a}}{a}(1 - 3w_i) + \frac{\dot{w_i}}{1+ w_i} \right) \theta_i 
     \nonumber \\ &\hspace{2.7cm} +\frac{\delta P_i/ \delta \rho_i}{1+ w_i} k^2 \delta_i - k^2 \sigma + k^2 \psi, \label{eqn:vel_per}
\end{align}
where the dot represents the derivative with respect to conformal time, $\tau$, which is related to the physical time $t$ through the relation $dt = ad\tau$, with $a$ being the scale factor. The variables $k$, $\sigma$, $\delta_i \equiv \delta \rho_i / \rho_i$, $\theta_i$ and $w_i$, denote the wavenumber of the Fourier mode, shear stress, density fluctuation, divergence of fluid velocity, and the EoS of the considered fluids, respectively. 

Furthermore, in the absence of anisotropic shear stress ($\sigma = 0$), Einstein's gravity imposes the condition that $\phi$ and $\psi$ are equal. Parallel to that, $\delta P_i/ \delta \rho_i $ (or $c^2_{a,i}$, the adiabatic sound speed of $i$-th fluid) is related to $w_i$ as 
\begin{eqnarray}
    c^{2}_{a,i} \equiv \frac{\delta P_i}{\delta \rho_i} = w_i - \frac{\dot{w_i}}{3\mathcal{H}(1+w_i)},
\end{eqnarray}
with $\mathcal{H}= \dot{a}/a$ being the conformal Hubble parameter, related to the physical Hubble parameter by $\mathcal{H}=aH$. It is important to note that equations \eqref{eqn:den_per} and \eqref{eqn:vel_per} apply both to a single uncoupled fluid and to the net (mass-averaged) quantities $\delta_i$ and $\theta_i$ across all fluids.
In this context, the Poisson equation in the  $k$-space is given by the relation \cite{Lima:1996at}, 
\begin{eqnarray}
    k^2 \phi = -4\pi G a^2 (\delta \rho + 3\delta p), \label{eqn:Possion_eq} 
\end{eqnarray}
where $\delta \rho = \delta \rho_1 + \delta \rho_2 $ and $\delta p = \delta p_1 + \delta p_2$. For this work, we are using $\delta \rho_i = \rho_i \delta_i$ and $\delta p_i = c^{2}_{a,i}\delta \rho_i$, where $i = 1, 2$. Thus, we can rewrite the Poisson equation \eqref{eqn:Possion_eq} as

\begin{eqnarray}
    k^2 \phi = -\frac{3}{2} \mathcal{H}^2  [(1+ 3c^{2}_{a,1})\Omega_1\delta_1 + (1+ 3c^{2}_{a,2})\Omega_2\delta_2], \label{eq:Poisson_mod}
\end{eqnarray}
where 
\begin{eqnarray}
    \Omega_i = \frac{\rho_i}{3H^2/8 \pi G} = \frac{\widetilde{\Omega}_i}{H^2/H^2_0} = \frac{\widetilde{\Omega}_i}{\sum_i \widetilde{\Omega}_i}
\end{eqnarray}
In our analysis, we assume that the EoS for both fluids remains constant ($w_i = \mathrm{constant}$ or $\dot{w_i} = 0$), which implies that the adiabatic sound speed satisfies $c^{2}_{a,i} = w_i$. Since the structure formation occurred in the matter-dominated regime, on spatial scales significantly smaller than horizon ($k \gg \mathcal{H}$), and with a negligible time variation of the gravitational potential $\phi$, we can disregard terms involving the derivative of $\phi$~\cite{Weller:2003hw, Unnikrishnan:2008qe, Jassal:2009ya, Jassal:2009gc}. Additionally, due to the negligible effect of dark energy perturbations on subhorizon scales, the second term in Eq.~\eqref{eq:Poisson_mod} can be omitted \cite{Peebles:1994xt, Abramo:2008ip, Creminelli:2008wc, Sapone:2009mb, Duniya:2013eta}. 

Under the assumed conditions, the modified evolution equations for the fluid $\rho_1$, can be expressed using Eqs.~\eqref{eqn:den_per}, \eqref{eqn:vel_per}, and \eqref{eq:Poisson_mod} as follows:

\begin{eqnarray}
    \dot{\delta_1} &=& -(1+w_1)\theta_1,\label{eqn:den_per_mod}\\
    \dot{\theta_1} &=& -\mathcal{H}(1 - 3w_1)\theta_1 + \frac{w_1}{1+w_1} k^2 \delta_1 \nonumber \\
    & & \hspace{2.2cm} - \frac{3}{2}\mathcal{H}^2 (1+3w_1)\Omega_1 \delta_1.
\end{eqnarray}

Moreover, using the relations

\begin{eqnarray}
\frac{d}{d\tau} = a \mathcal{H} \frac{d}{da}; \quad 
\frac{d^2}{d\tau^2} = \left(a\mathcal{H}^2 + a^2 \mathcal{H} \frac{d\mathcal{H}}{da} \right) \frac{d}{da} + a^2 \mathcal{H}^2 \frac{d^2}{da^2}, \nonumber
\end{eqnarray}
we can write the second-order differential equation governing the perturbation of the fluid $\rho_1$ as

\begin{align}
    &\delta_1^{\prime \prime} + \bigg(\frac{\mathcal{H}(2-3w_1) + a\mathcal{H}^{\prime}}{a \mathcal{H}}\bigg)\delta^{\prime}_{1} + \nonumber \\
   &\hspace{0.7cm} \bigg(\frac{w_1 k^2 - (3/2) \mathcal{H}^2 (1+ w_1)(1+3w_1)\Omega_1}{a^2 \mathcal{H}^2}\bigg)\delta_1 = 0,\label{eqn:delta_non_inter}
\end{align}
where the prime represents the derivative with respect to the scale factor $a$.

\subsubsection{Interacting scenario}

For interacting scenarios, where energy transfer occurs between the two fluids, characterized by $g(a)$ as given in Eq. \eqref{eqn-fa},  the evolution of perturbations in Fourier space $k$ within the conformal Newtonian gauge follows from the perturbed conservation equations, incorporating the effects of the perturbed metric and energy-momentum tensor \cite{Marcondes:2016reb}

\begin{align}
    \dot{\delta}_i &= -\bigg(3\mathcal{H}(c^2_{a,i} - w_i ) - \frac{Q_0}{\rho_i}  \bigg)\delta_i -  \nonumber \\
    & \hspace{3.2cm}  (1+w_i)(\theta_i - 3\dot{\phi}) - \frac{\delta Q_0}{\rho_i},   \label{eqn:int_den_per} \\
    \dot{\theta}_i &= -\bigg(\mathcal{H}(1-3w_i) - \frac{Q_0}{\rho_i} + \frac{\dot{w_i}}{1+w_i} \bigg)\theta_i + \nonumber \\
    & \hspace{2.4cm} k^2\phi + \frac{c^2_{a, i}}{1+w_i}k^2\delta_i + \frac{ik^j \delta Q_j}{\rho_i(1+w_i)}. \label{eqn:int_vel_per}
\end{align}
Here, we have neglected the anisotropic stress, assuming $\phi = \psi$. The variables $k$, $\sigma$, $\delta_i$, $\theta_i$ and $w_i$ are the same ones as discussed in Sec. \ref{subsec:non-int_per}.

The term $Q$ represents the interaction function, with $Q_0$ corresponding to its temporal component and $Q_j$ denoting the spatial part. In our analysis, we consider only the temporal component where  $Q_0$ is given by Eq.~\eqref{eqn:inter_func_model}, while the spatial part $Q_j$ is set to zero. Consequently, the perturbation to the exchange of energy-momentum in the conservation equation is given by:
\begin{eqnarray}
    \delta Q = \delta Q_0 = \frac{\delta H}{2}(\xi_1 \rho_1 + \xi_2 \rho_2) + \nonumber \\
     \frac{H}{2}(\xi_1 \delta \rho_1 + \xi_2 \delta \rho_2).  \label{eqn:delta_Q}
\end{eqnarray}
Now, since the impact of $\delta H$ on the evolution of matter density is quantitatively negligible \cite{Gavela:2010tm, Yang:2014hea}, we have therefore neglected the perturbation of $H$ in our analysis. 

Considering the conditions as discussed in Sec.  \ref{subsec:non-int_per} and using Eqs. \eqref{eq:Poisson_mod} and \eqref{eqn:delta_Q}, we can  rewrite Eqs. \eqref{eqn:int_den_per} and \eqref{eqn:int_vel_per} for the fluid $\rho_1$ as 
\begin{eqnarray}
    \dot{\delta}_1 &=& -\bigg( \frac{\mathcal{H}}{2a}\xi_1 -\frac{Q}{\rho_1} \bigg)\delta_1 - (1+w_1)\theta_1, \\
    \dot{\theta_1} &=& -\bigg(\mathcal{H}(1-3w_1) - \frac{Q}{\rho_1}\bigg)\theta_1 + \frac{w_1}{1+w_1} k^2 \delta_1 - \nonumber\\ 
    & &\hspace{3.1cm} \frac{3}{2} \mathcal{H}^2 (1+3w_1)\Omega_1 \delta_1.
\end{eqnarray}
In terms of the scale factor $a$, the evolution of fluid perturbations can be described by the second-order differential equation as

\begin{widetext}
\begin{align}
    &\delta^{\prime \prime}_1 + \bigg[ a\mathcal{H}^{\prime} + \bigg(2 - 3w_1 + \frac{\xi_1}{2a} \bigg) \mathcal{H}  -\frac{2Q}{\rho_1}  \bigg]\frac{\delta_1^{\prime}}{a\mathcal{H}} + \bigg[\bigg((a\mathcal{H}^\prime - 3 w_1\mathcal{H})\rho_1   - Q \bigg)\frac{\mathcal{H} \xi_1}{2a \rho_1} + \nonumber \\
    &\hspace{2.7cm} \bigg( \frac{Q^{2} +  a \mathcal{H}(Q \rho^{\prime}_1 - Q^{\prime} \rho_1) - Q\rho_{1}\mathcal{H}(1 - 3w_1)}{\rho^{2}_{1}} \bigg) + 
    w_{1}k^{2} - \frac{3}{2} \mathcal{H}^2 (1+w_1)(1+3w_1)\Omega_1   \bigg]\frac{\delta}{a^2 \mathcal{H}^2} = 0. \label{eqn:delta_inter}
\end{align}
\end{widetext}

To estimate the density contrast $\delta_{1}$ as a function of the scale factor, we numerically integrate Eqs. \eqref{eqn:delta_non_inter} and \eqref{eqn:delta_inter} from an initial scale factor of $a = 10^{-3}$ up to  $a =1$ (present epoch), adopting the initial conditions as 
outlined in Ref. \cite{Abramo:2008ip},
\begin{eqnarray}
    \delta_{1,j} = -2\phi_j \bigg(1 + \frac{k^2}{3\mathcal{H}_j} \bigg), \quad \delta^{\prime}_{1,j} = -\frac{2}{3}\frac{k^2}{\mathcal{H}_j}\phi_j
\end{eqnarray}
where $j$ represents the values of respective parameters at $a_j = 10^{-3}$. Although the choice of initial conditions does not impact the robustness of the density contrast solution, for consistency, we adopt $\phi_j = -6 \times 10^{-7}$ in this work \cite{Mehrabi:2015hva}. This choice ensures that the dark matter density contrast reaches $0.1$ at the present epoch for $k=0.1h\mathrm{Mpc}^{-1}$. Moreover, selecting $k=0.1h\mathrm{Mpc}^{-1}$ aligns well with the power-spectrum results from Galaxy surveys for scales $k \leq 0.15h\mathrm{Mpc}^{-1}$  and remains consistent with the power-spectrum normalization $\sigma_8$, which corresponds to $k = 0.125h\mathrm{Mpc}^{-1}$ \cite{Smith:2002dz, SDSS:2003eyi, Percival:2006gt}.

To compare the analytical solution of density fluctuation of the fluid $\rho_1$ with current observation, we calculate a quantity $f\sigma_8(a)$, which is given by the relation~\cite{Nesseris:2007pa}
\begin{align}
    f\sigma_8(a) &\equiv f(a) \cdot \sigma_8(a)\nonumber \\
    &= \sigma_{8,0} \frac{a}{\delta_i(a=1)} \frac{d \delta_i(a)}{da},  
\end{align}
where $f(a)$ represents the linear growth of matter perturbation as a function of the scale factor $a$, which quantifies the formation and evolution of cosmic structures. On the other hand,  $\sigma_8$ denotes the root-mean-square fluctuations of the linear density field in  a sphere with  a radius  $R = 8 h^{-1}\mathrm{Mpc}$. The relation of $f(a)$ and $\sigma_8(a)$ is expressed as~\cite{Nesseris:2007pa}
\begin{align}
    f(a) = \frac{d~ln\delta_i}{d~lna} = \frac{a}{\delta_i(a)}\frac{d\delta_i}{da}~;\quad \sigma_8 = \sigma_{8,0} \frac{\delta_i(a)}{\delta_i(1)}.
\end{align}

Here, for our analysis, we chose a prior on $\sigma_{8,0} = 0.8111 \pm 0.0060$ at 68\% confidence level,  as determined by the Planck 2018 results ~\cite{Planck:2018vyg}. A detailed comparison is presented in Sec.~\ref{sec-cons} between the predictions of the prescribed model -- constrained using various observational datasets employed in this study -- and those of the $\Lambda$CDM model constrained by Planck 2018 results~\cite{Planck:2018vyg}, with both being evaluated against recent $f\sigma_8$ measurements, as summarized in Table \ref{tab:fsigma8}.

\section{Observational data and Methodology}
\label{data}

\begin{table}[t!]
\begin{center}
\resizebox{0.48\textwidth}{!}{%
    \huge
\renewcommand{\arraystretch}{0.1} 
\begin{tabular}{ | @{\hspace{0.1cm}} l @{\hspace{0.4 cm}} l @{\hspace{0.4 cm}} l @{\hspace{0.3 cm}} c @{\hspace{0.3 cm}} c @{\hspace{0.1 cm}} |@{\hspace{0.3cm}} l @{\hspace{0.3 cm}} l @{\hspace{0.4 cm}} l @{\hspace{0.4 cm}} c @{\hspace{0.3 cm}} c @{\hspace{0.1 cm}}|}
\multicolumn{10}{c}{{}}\\
\hline 
\hline
~$z$ & $H(z)$  & $\sigma_{H(z)}$ & Method & Ref. & $z$ & $H(z)$  & $\sigma_{H(z)}$ & Method & Ref. \\
\hline
~0.07  & 69.0  & 19.6  & DA & \cite{Zhang:2012mp} & 0.51  & 90.4  & 1.9   & Clustering & \cite{BOSS:2016wmc} \\
~0.1   & 69.0  & 12.0  & DA  & \cite{Stern:2009ep} & 0.52  & 94.35 & 2.65  & Clustering & \cite{BOSS:2016zkm} \\
~0.12  & 68.6  & 26.2  & DA & \cite{Zhang:2012mp} & 0.56  & 93.33 & 2.32  & Clustering & \cite{BOSS:2016zkm} \\
~0.17  & 83.0  & 8.0   & DA & \cite{Stern:2009ep} & 0.57  & 92.9  & 7.8   & Clustering & \cite{BOSS:2013kxl} \\
~0.1797 & 75.0  & 4.0   & DA & \cite{Moresco:2012jh} & 0.59  & 98.48 & 3.19  & Clustering & \cite{BOSS:2016zkm} \\
~0.1993 & 75.0  & 5.0   & DA & \cite{Moresco:2012jh} & 0.5929 & 104.0 & 13.0  & DA & \cite{Moresco:2012jh} \\
~0.20  & 72.9  & 29.6  & DA  & \cite{Zhang:2012mp} & 0.6   & 87.9  & 6.1   & Clustering & \cite{Blake:2012pj} \\
~0.24  & 79.69 & 2.65  & Clustering & \cite{Gaztanaga:2008xz} & 0.61  & 97.3  & 2.1   & Clustering & \cite{BOSS:2016wmc} \\
~0.27  & 77.0  & 14.0  & DA & \cite{Stern:2009ep} & 0.64  & 98.82 & 2.99  & Clustering & \cite{BOSS:2016zkm} \\
~0.28  & 88.8  & 36.6  & DA & \cite{Zhang:2012mp} & 0.6797 & 92.0  & 8.0   & DA & \cite{Moresco:2012jh} \\
~0.3   & 81.7  & 6.22  & Clustering & \cite{Oka:2013cba} & 0.73  & 97.3  & 7.0   & Clustering & \cite{Blake:2012pj} \\
~0.31  & 78.17 & 4.74  & Clustering & \cite{BOSS:2016zkm} & 0.75  & 98.8  & 33.6  & Clustering & \cite{Borghi:2021rft} \\
~0.34  & 83.8  & 3.66  & Clustering & \cite{Gaztanaga:2008xz} & 0.7812 & 105.0 & 12.0  & DA & \cite{Moresco:2012jh} \\
~0.35  & 82.7  & 8.4   & Clustering & \cite{Chuang:2012qt} & 0.8754 & 125.0 & 17.0  & DA & \cite{Moresco:2012jh} \\
~0.3519 & 83.0  & 14.0  & DA & \cite{Moresco:2012jh} & 0.88  & 90.0  & 40.0  & DA & \cite{Stern:2009ep} \\
~0.36  & 79.93 & 3.39  & Clustering & \cite{BOSS:2016zkm} & 0.9   & 117.0 & 23.0  & DA & \cite{Stern:2009ep} \\
~0.38  & 81.5  & 1.9   & Clustering & \cite{BOSS:2016wmc} & 0.978 & 113.72 & 14.63 & Clustering & \cite{eBOSS:2018yfg} \\
~0.3802 & 83.0  & 13.5  & DA & \cite{Moresco:2016mzx} & 1.037 & 154.0 & 20.0  & DA & \cite{Moresco:2012jh} \\
~0.40  & 82.04 & 2.03  & DA & \cite{BOSS:2016zkm} & 1.23  & 131.44 & 12.42 & Clustering & \cite{eBOSS:2018yfg} \\
~0.4   & 95.0  & 17.0  & DA & \cite{Stern:2009ep} & 1.3   & 168.0 & 17.0  & DA & \cite{Stern:2009ep} \\
~0.4004 & 77.0  & 10.2  & DA & \cite{Moresco:2016mzx} & 1.363 & 160.0 & 33.6  & DA & \cite{Moresco:2015cya} \\
~0.4247 & 87.1  & 11.2  & DA & \cite{Moresco:2016mzx} & 1.43  & 177.0 & 18.0  & DA & \cite{Stern:2009ep} \\
~0.4293 & 91.8  & 5.3   & DA & \cite{Moresco:2016mzx} & 1.526 & 148.11 & 12.71 & Clustering & \cite{eBOSS:2018yfg} \\
~0.43  & 86.45 & 3.68  & Clustering & \cite{Gaztanaga:2008xz} & 1.53  & 140.0 & 14.0  & DA & \cite{Stern:2009ep} \\
~0.44  & 82.6  & 7.8   & Clustering & \cite{Blake:2012pj} & 1.75  & 202.0 & 40.0  & DA & \cite{Stern:2009ep} \\
~0.44  & 84.81 & 1.83  & Clustering & \cite{BOSS:2016zkm} & 1.944 & 172.63 & 14.79 & Clustering & \cite{eBOSS:2018yfg} \\
~0.4497 & 92.8  & 12.9  & DA & \cite{Moresco:2016mzx} & 1.965 & 186.5 & 50.4  & DA & \cite{Moresco:2015cya} \\
~0.47  & 89.0  & 34.0  & DA & \cite{Ratsimbazafy_2017} & 2.3   & 224.0 & 8.0   & Clustering & \cite{BOSS:2014hwf} \\
~0.4783 & 80.9  & 9.0   & DA & \cite{Moresco:2016mzx} & 2.33  & 224.0 & 8.0   & Clustering & \cite{BOSS:2017fdr} \\
~0.48  & 87.79 & 2.03  & DA & \cite{BOSS:2016zkm} & 2.34  & 222.0 & 7.0   & Clustering & \cite{BOSS:2014hwf} \\
~0.48  & 97.0  & 62.0  & DA & \cite{Stern:2009ep} & 2.36  & 226.0 & 8.0   & Clustering & \cite{BOSS:2013igd} \\
\hline

\end{tabular}}
\label{table_OHD}
\end{center}
\caption{The complete set of 62 observational Hubble Data (OHD), along with their associated uncertainties, utilized in this work. Here, {\bf DA} refers to the differential age (cosmic chronometer) method, while ``clustering'' corresponds to measurements from radial BAO surveys. }
\label{table-hubble}
\end{table}

This section explores the cosmological probes from various observational surveys and the statistical methodology employed in this work to constrain the parameters associated with the proposed dark energy model discussed in the previous section. The cosmological probes are outlined as follows:

\begin{itemize}
    \item {\bf Observational Hubble Data (OHD)}: The Hubble parameter $H(z)$ plays a crucial role in constraining a DE model, particularly in the low-redshift regime. One of the two approaches to measure $H(z)$ involves determining the differential age (DA) between two groups of passively evolving galaxies at varying redshift,  often known as cosmic chronometers. Another alternative approach is to examine the radial (line-of-sight) scale of BAO features at different redshifts, known as the radial BAO size method. In this work, we utilize a total of 62 OHD data points \cite{Yang:2023qsz}, comprising 34 data points  that are obtained through the DA method and 28 data points derived from the radial BAO size method, as shown in Table~\ref{table-hubble}.

    \item {\bf Baryon Acoustic Oscillation:} We also utilize measurements from BAO, specifically from DESI and SDSS, where BAO matter clustering is treated as a standard ruler. In both cases, isotropic BAO measurements are expressed as  $D_V(z)/r_d$, representing the angle-averaged distance normalized to the comoving sound horizon at the drag epoch. In contrast, anisotropic BAO measurements include $D_M(z)/r_d$ and $D_H(z)/r_d$, where $D_M$ is the comoving angular diameter distance, and $D_H$ is the Hubble horizon. Below, we briefly describe the data from each experiment: 
    \begin{itemize}
        \item[1.] {\bf Dark Energy Spectroscopic Instrument}: 
       The year-1 and year-2 data releases include tracers from the Bright Galaxy Sample (BGS), Luminous Red Galaxies (LRG), Emission Line Galaxies (ELG), Quasars (QSO), and the Lyman-$\alpha$ forest. In this article, we utilize data from both DESI datasets, summarized as follows:\\ 
       
       (a) \textbf{DESI Data Release 1 (DESI-DR1)} consists of both isotropic and anisotropic measurements across the redshift range $0.1 < z < 4.2$, as outlined in Table I of Ref.~\cite{DESI:2024mwx}. We refer to this dataset as {\bf DESI-DR1}. 
       
       (b) \textbf{DESI Data Release 2 (DESI-DR2)} builds upon the earlier release by incorporating measurements spanning the redshift range $0.0295 \leq z \leq 2.330$, divided into nine redshift bins, as detailed in Table IV of Ref.~\cite{DESI:2025zgx}. Throughout this work, we denote this dataset as {\bf DESI-DR2}.

        \item[2.] {\bf Sloan Digital Sky Survey:}  Spectroscopic galaxy and quasar samples from four generations of SDSS include the Main Galaxy Sample (MGS) ($0.07 < z < 0.2$), BOSS DR12 galaxies ($0.2 < z < 0.6$), eBOSS galaxies and Quasars ($0.6 < z < 2.2$), and Lyman-$\alpha$ forest-samples ($1.8 < z < 3.5$). In our analysis, we specifically use BAO-only measurements from Table III of Ref.~\cite{eBOSS:2020yzd}. We denote this dataset as {\bf SDSS}. 
    \end{itemize} 
    Accordingly, when incorporating BAO measurements into our analysis, we adopt the value of $r_d = 147.09 \pm 0.26 ~\mathrm{Mpc} $  as determined in Ref.~\cite{Planck:2018vyg}.

    \item {\bf Type Ia Supernovae  (SNIa):} The discovery of the accelerated expansion of the Universe was first revealed through observations of  SNIa. Their widespread use as standard candles arises from their relatively uniform absolute luminosity. In this study, we incorporate three observational datasets derived from SNIa, detailed as follows:
    \begin{itemize}
       \item[1.] {\bf Pantheon$+$:} Pantheon$+$, the successor to the Pantheon sample, comprises 1701 light-curve measurements  of 1550 distinct supernovae, distributed over the redshift interval $z = [0.001, 2.26]$~\cite{Scolnic:2021amr, Brout:2022vxf}. In this article, we consider the subset of 1590 SNIa data points,  within the redshift range $0.01016 \leq z \leq 2.26137$, in order to reduce the impact of the peculiar velocity corrections ~\cite{Chan-GyungPark:2024mlx}. We refer to this dataset as {\bf Pan$+$}. 
    
       \item[2.] {\bf Union3:} The up-to-date Union3 compilation consists  of 2087 SNIa from 24 datasets, which are standardized to a consistent distance scale using SALT3 light-curve fitting. These data are subsequently analyzed and binned using the UNITY1.5 Bayesian framework, resulting in 22 binned distance modulus measurements in the redshift range $0.05<z<2.26$, which we adopt in this article~\cite{Rubin:2023ovl}.  
       
       \item[3.] {\bf DESY5:} We utilize the full five year dataset from the Dark Energy Survey supernova program, comprising 1635 newly photometrically classified SNIa within the redshift range $0.1< z < 1.3$. This dataset is further supplemented by 194 low-redshift SNIa from CfA3~\cite{Hicken:2009df}, CfA4~\cite{Hicken:2012zr}, Carnegie
Supernova Project \cite{Krisciunas:2017yoe}, and Foundation~\cite{Foley:2017zdq} samples, covering the redshift range $0.025 < z < 0.1$~\cite{DES:2024jxu}.
     \end{itemize}
\end{itemize}

In order to fit our model using various observational datasets, we perform the Markov chain Monte Carlo (MCMC) analysis employing an affine-invariant \texttt{\small PYTHON}-based sampler \textbf{\texttt{\small EMCEE}} \cite{Foreman-Mackey:2012any} which is publicly available.\footnote{\href{https://github.com/dfm/emcee}{https://github.com/dfm/emcee}} We adopt the uniform priors on the model parameters as outlined: $H_0 \thicksim \mathcal{U}(50, 100)$, $\Omega_{0} \thicksim \mathcal{U}(0,1)$, $\delta \thicksim \mathcal{U}(-5,5)$ and $\alpha \thicksim \mathcal{U}(-5,5)$. For the analysis and visualization of the MCMC chains, we employ the widely used package \textbf{\texttt{\small GETDIST}}~\cite{Lewis:2019xzd}.\footnote{\href{https://github.com/cmbant/getdist}{https://github.com/cmbant/getdist}}

\section{Results}
\label{sec-results}

In this section, we present the outcomes of our model, focusing on its observational constraints, thermodynamical implications, and its fitness with the observational data, quantified by the model comparison statistics. 

\subsection{Observational analysis for parameter estimation}
\label{sec-cons}

\begin{table*}
        \centering
\resizebox{1.0 \textwidth}{!}{%
    \huge 
\renewcommand{\arraystretch}{1.8} 
        \begin{tabular}{l c c c c}
           \hline
              ~Datasets ~ & ~$H_0$~  & ~$\Omega_0$~ &  ~$\delta$~ & ~$\alpha$~ \\
            \hline
            \hline

                ~OHD+Pan$+$~ & ~$67.02^{+0.85}_{-0.85}\left(67.0^{+1.7}_{-1.7}\right) $~ & ~$0.458^{+0.170}_{-0.097}\left(0.46^{+0.23}_{-0.28}\right) $~ & ~$-0.35^{+0.11(+0.56)}_{-0.31(-0.40)} $~ & ~$-0.61^{+1.10}_{-0.60}\left( -0.6^{+1.6}_{-1.9}\right) $~ \\
                
                ~OHD+DESI-DR1+Pan$+$~ & ~$67.82^{+0.69}_{-0.69}\left(67.4^{+1.4}_{-1.3} \right) $~ & ~$0.35^{+0.11(+0.20)}_{-0.11(-0.21)} $~ & ~$-0.11^{+0.15(+0.50)}_{-0.30(-0.42)} $~ & ~$-0.01^{+0.63(+0.95)}_{-0.38(-1.10)} $~ \\

                ~OHD+DESI-DR2+Pan$+$~ & ~$67.84^{+0.59}_{-0.59}\left(67.8^{+1.2}_{-1.2}\right) $~ & ~$0.333^{+0.081}_{-0.081} \left(0.33^{+0.15}_{-0.16} \right)  $~ & ~$-0.08^{+0.13(+0.39)}_{-0.22(-0.34)} $~ & ~$0.07^{+0.46(+0.77)}_{-0.34(-0.82)} $~ \\

                ~OHD+SDSS+Pan$+$~ & ~$67.15^{+0.67}_{-0.67}\left( 67.2^{+1.3}_{-1.3} \right) $~ & ~$0.479^{+0.140}_{-0.076} \left(0.48^{+0.20}_{-0.23} \right) $~ & ~$-0.384^{+0.092}_{-0.240} \left(-0.38^{+0.43}_{-0.33} \right) $~ & ~$-0.72^{+0.96}_{-0.65}\left(-0.7^{+1.5}_{-1.7} \right) $~ \\[0.5pt]

                \multicolumn{5}{ c }{\hdashrule{\dimexpr\textwidth-2\tabcolsep\relax}{0.5pt}{3mm 2mm}}\\

                ~OHD+Union3~ & ~$66.50^{+1.1(+2.2)}_{-1.1(-2.1)}$~ & ~$0.37^{+0.20(+0.28)}_{-0.16(-0.31)}$~ & ~$-0.14^{+0.16(+0.87)}_{-0.49(-0.60)}$~ & ~$0.03^{+1.10(+1.4)}_{-0.43(-1.8)}$~ \\

                ~OHD+DESI-DR1+Union3~ & ~$67.37^{+0.90(+1.8)}_{-0.90(-1.7)}$~ & ~$0.29^{+0.12(+0.21)}_{-0.13(-0.22)}$~ & ~$0.05^{+0.19(+0.68)}_{-0.40(-0.55)}$~ & ~$0.33^{+0.65(+0.94)}_{-0.36(-1.10)}$~ \\

                ~OHD+DESI-DR2+Union3~ & ~$67.33^{+0.78}_{-0.78}\left(67.3^{+1.6}_{-1.5} \right)$~ & ~$0.275^{+0.094}_{-0.094}\left(0.28^{+0.18}_{-0.18} \right) $~ & ~$0.08^{+0.17(+0.56)}_{-0.33(-0.47)} $~ & ~$0.40^{+0.53(+0.82)}_{-0.34(-0.90)} $~ \\

                ~OHD+SDSS+Union3~ & ~$66.65^{+0.9(+1.8)}_{-0.9(-1.7)}$~ & ~$0.41^{+0.18(+0.25)}_{-0.11(-0.29)}$~ & ~$-0.23^{+0.12(+0.66)}_{-0.37(-0.47)}$~ & ~$-0.16^{+1.10(+1.40)}_{-0.54(-1.70)}$~ \\

                \multicolumn{5}{ c }{\hdashrule{\dimexpr\textwidth-2\tabcolsep\relax}{0.5pt}{3mm 2mm}}\\
            
                ~OHD+DESY5~ & ~$69.67^{+0.37(+0.73)}_{-0.37(-0.70)}$~ & ~$0.208^{+0.072}_{-0.160}\left(0.21^{+0.22}_{-0.19}\right)$~ & ~$0.23^{+0.27(+0.99)}_{-0.59(-0.80)}$~ & ~$0.56^{+0.53(+0.70)}_{-0.24(-0.86)}$~ \\

                ~OHD+DESI-DR1+DESY5~ & ~$69.70^{+0.36(+0.71)}_{-0.36(-0.69)}$~ & ~$0.158^{+0.05}_{-0.09}\left(0.16^{+0.14}_{-0.12}\right)$~ & ~$0.44^{+0.26(+0.72)}_{-0.42(-0.65)}$~ & ~$0.67^{+0.33(+0.50)}_{-0.20(-0.56)}$~ \\

                ~OHD+DESI-DR2+DESY5~ & ~$69.75^{+0.35(+0.71)}_{-0.35(-0.68)} $~ & ~$0.182^{+0.050}_{-0.067}\left(0.18^{+0.11}_{-0.11} \right)$~ & ~$0.33^{+0.19(+0.49)}_{-0.28(-0.45)} $~ & ~$0.54^{+0.29(+0.46)}_{-0.20(-0.50)} $~ \\
  
                ~OHD+SDSS+DESY5~ & ~$69.74^{+0.37(+0.74)}_{-0.37(-0.70)}$~ & ~$0.239^{+0.090}_{-0.13}\left(0.24^{+0.20}_{-0.19}\right)$~ & ~$0.10^{+0.22(+0.74)}_{-0.44(-0.61)}$~ & ~$0.40^{+0.50(+0.72)}_{-0.28(-0.82)}$~ \\
            \hline
            \hline
        \end{tabular}

    }
         \caption{ Marginalized constraints at 68\% and 98\% CL on the four model parameters, derived from various observational datasets.}
    
        \label{tab:Psi-mcmc}
    \end{table*}

\begin{figure}
   \centering
    \includegraphics[width=0.5\textwidth]{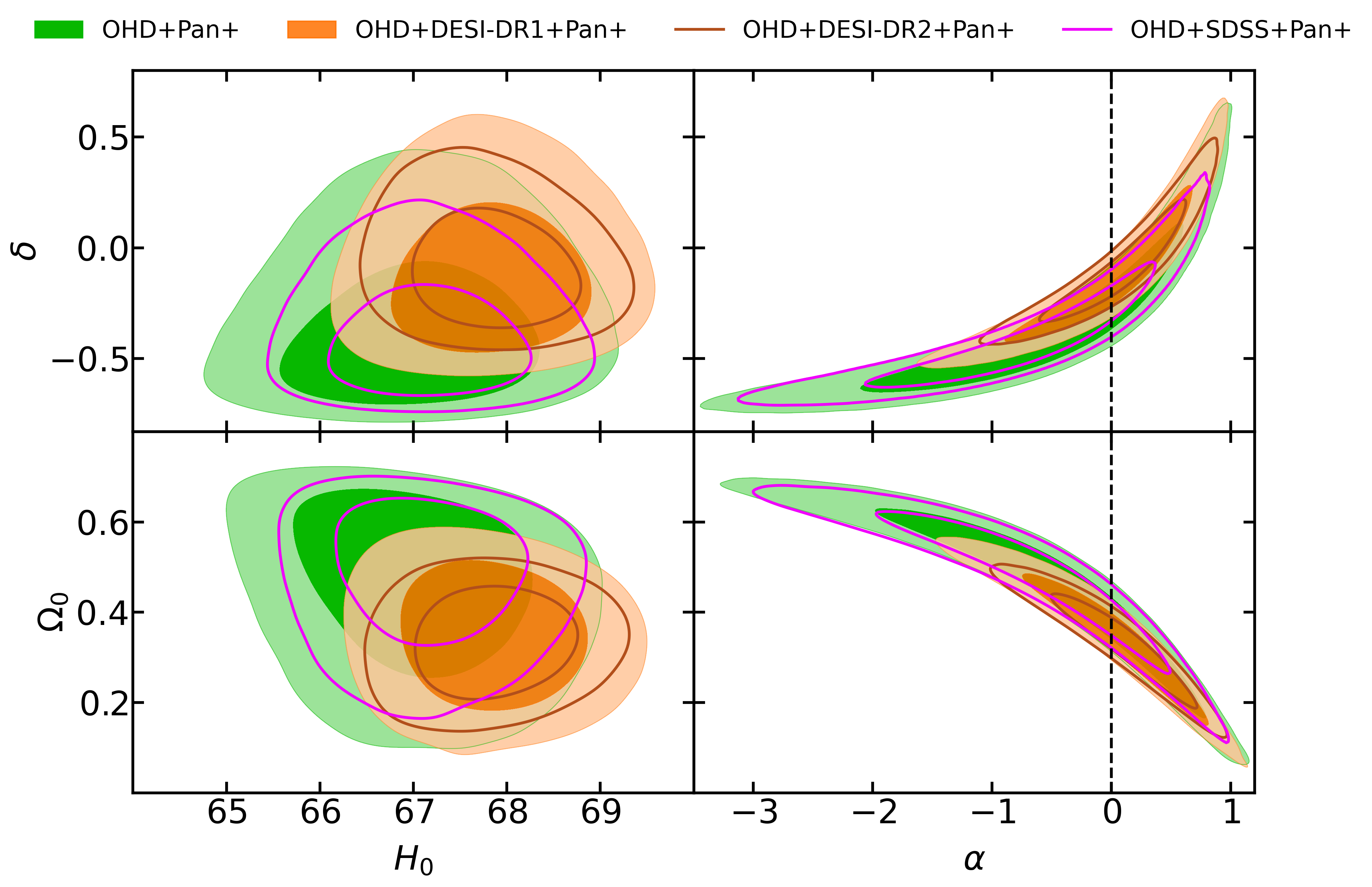}
    \includegraphics[width=0.5\textwidth]{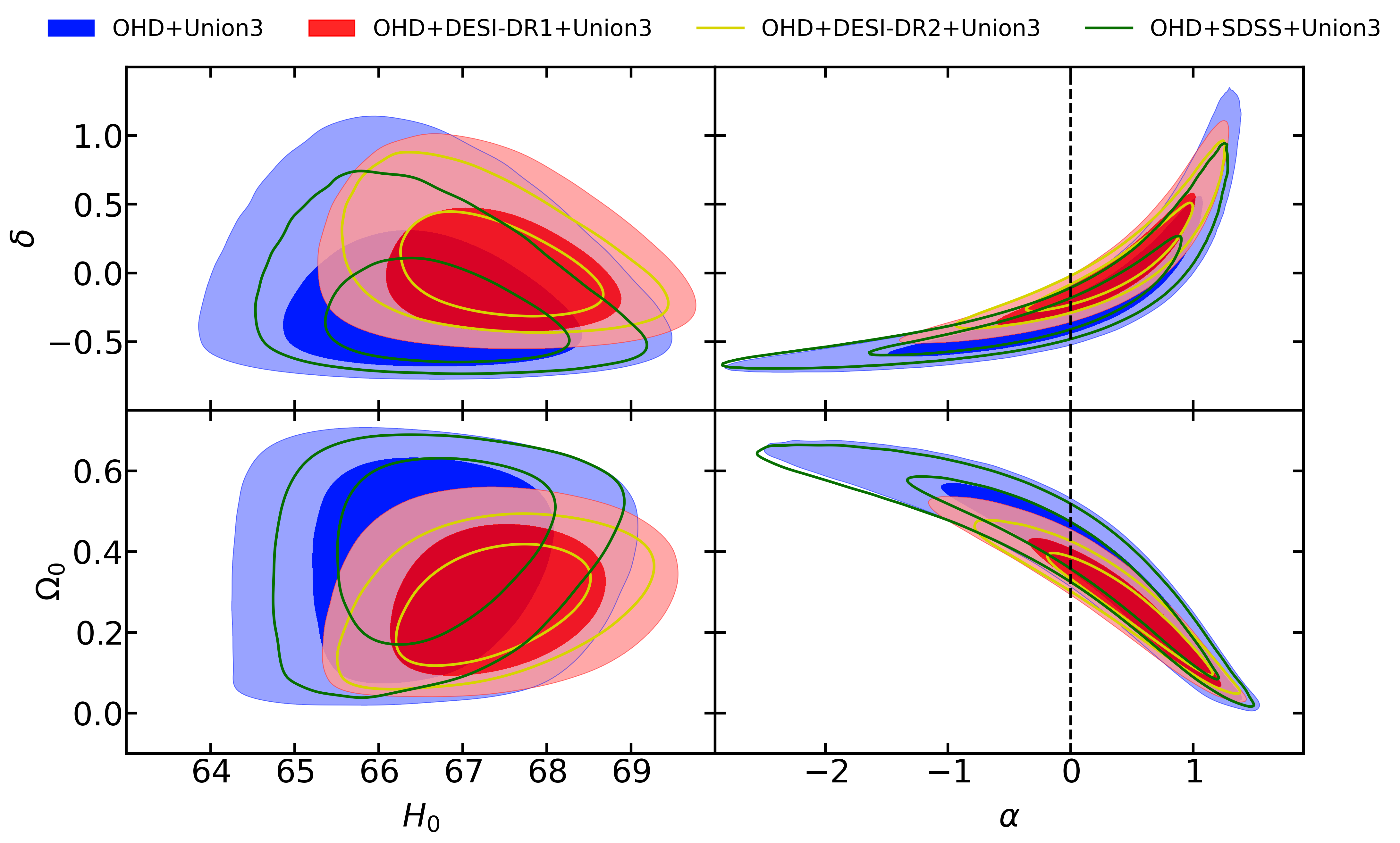}
    \includegraphics[width=0.5\textwidth]{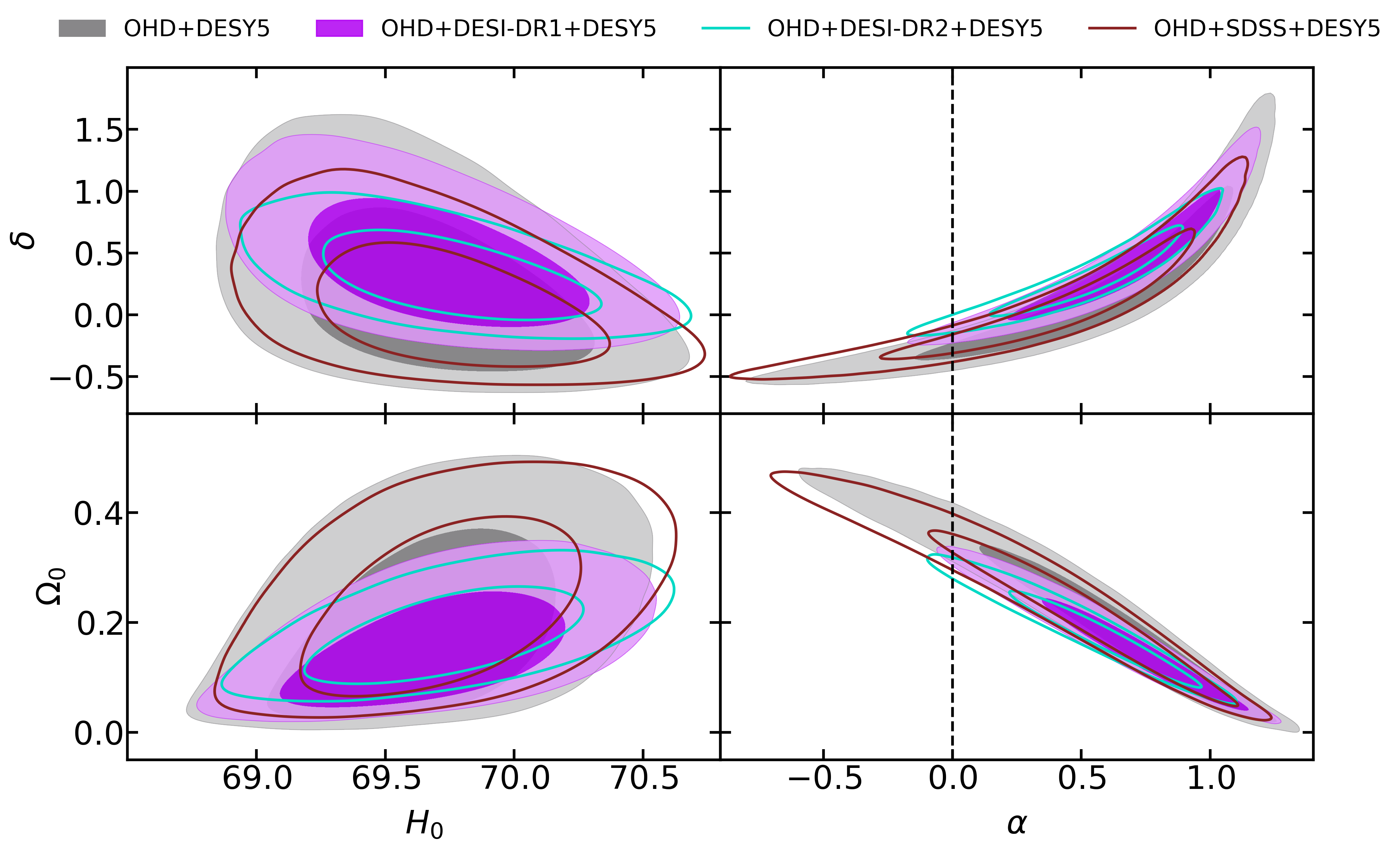}
    \caption{The marginalized 2D posterior distribution with 2D contours  (68\% and 95\% CL) for the cosmological parameters $H_0$, $\Omega_0$, $\delta$, and $\alpha$. The dotted lines in the 2D contours represent $\delta=0$ and $\alpha=0$. Results are shown for various datasets specified in the legend.} 
    \label{fig:contour}
\end{figure}
We now proceed to a detailed discussion of the results derived from the observational datasets described earlier. As mentioned in the previous section, we employed three SNIa datasets, namely, Pan$+$, Union3, and DESY5, and in each case, we aim to investigate the effects of the DESI (both DR1 and DR2) and SDSS datasets separately. The dataset related to the OHD measurements remains fixed across all combinations. Therefore, our primary focus is to test the results using the combination of OHD and SNIa data. In the next step, we include the BAO measurements, leading to three joint analyses: OHD + DESI-DR1 + SNIa, OHD + SDSS + SNIa, and OHD + DESI-DR2 + SNIa.

\begin{figure*}
    \includegraphics[width=1.02\textwidth]{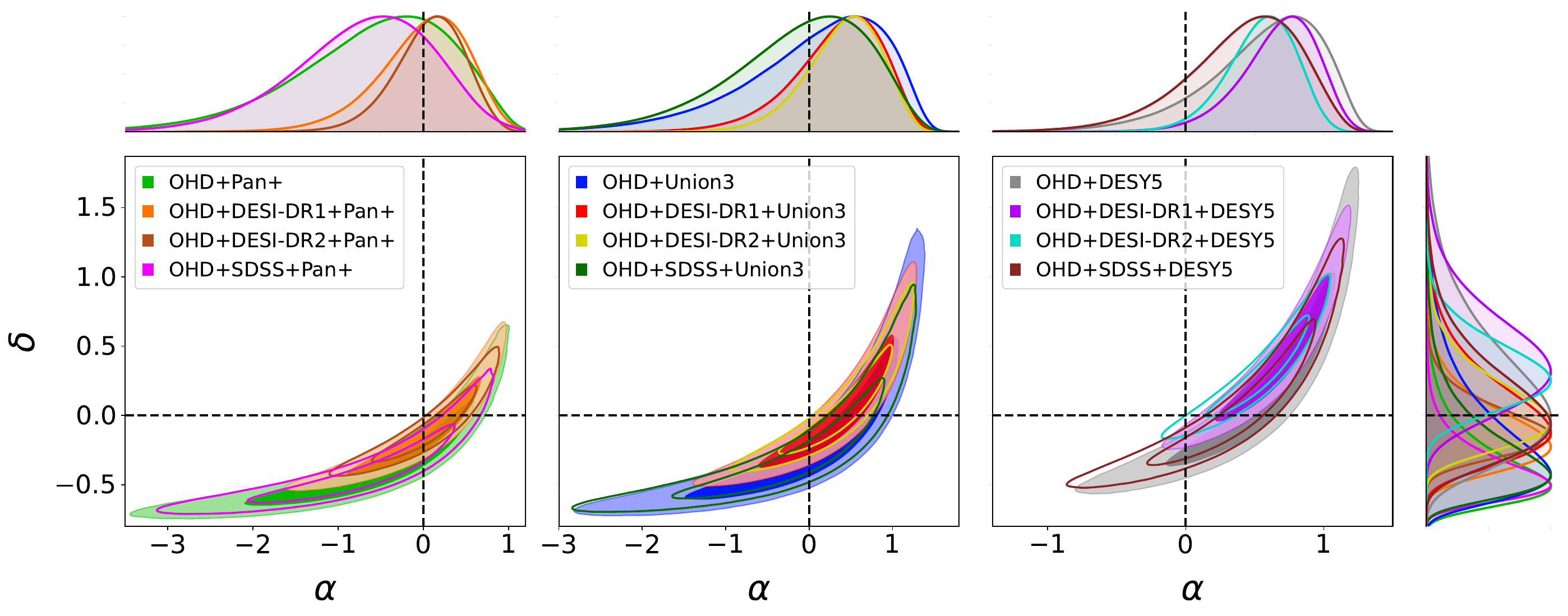}
    \caption{One-dimensional probability distribution functions and two-dimensional marginalized contours in the $\alpha-\delta$ plane for different dataset combinations, as described in Sec.~\ref{sec-cons}. The horizontal and vertical reference lines indicate the parameter values at which our model recovers the standard $\Lambda$CDM model. }
    \label{fig:delta-alpha}
\end{figure*}

\begin{figure}
    \centering
    \includegraphics[width=0.48\textwidth]{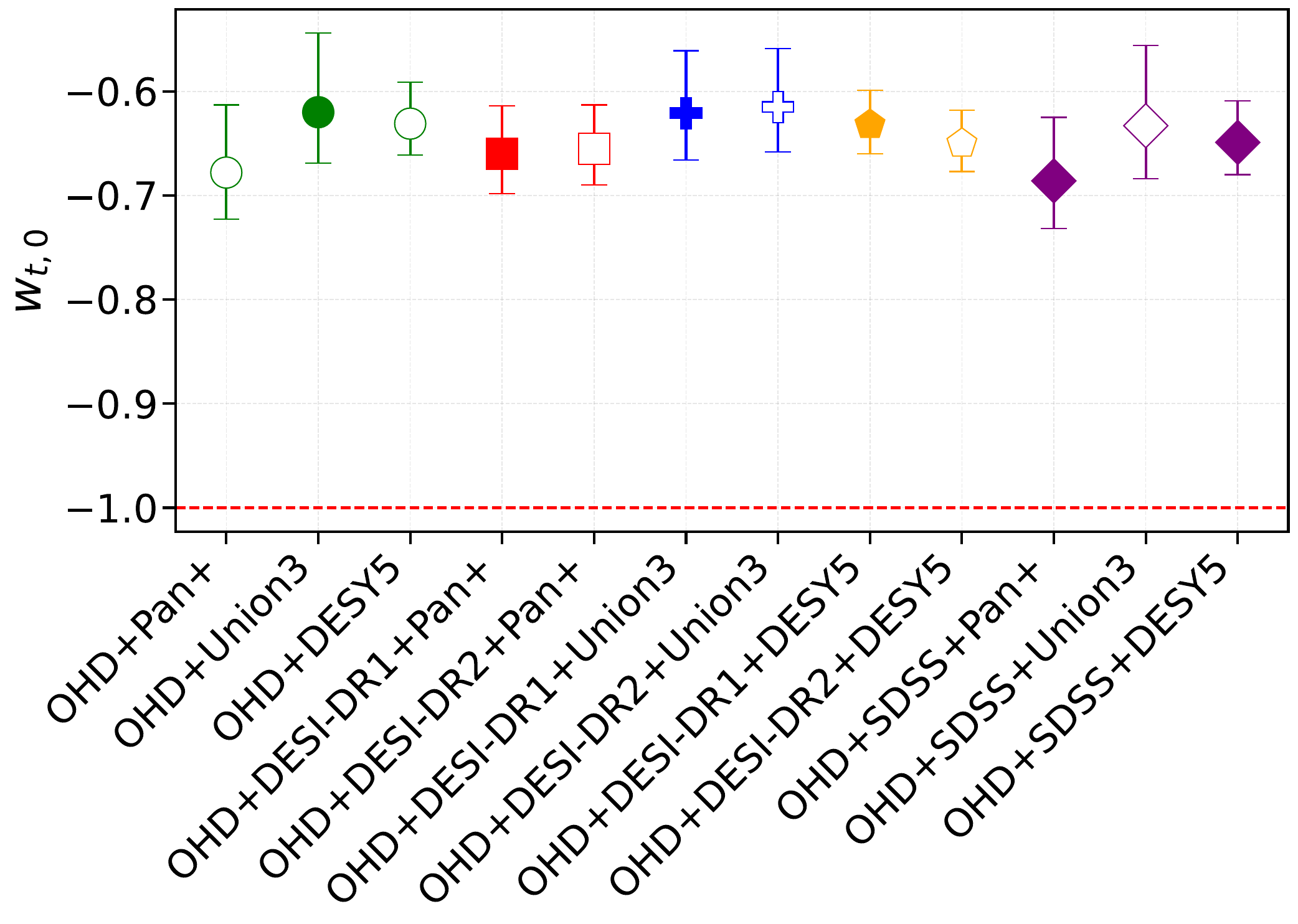}    \caption{ The present-day value of the effective EoS $w_{t,0}$, derived based on the observational constraints for all  considered dataset combinations.  The red dashed line denotes the phantom divide, separating the quintessential regime from the phantom regime.}
    \label{fig:Eos-today}
\end{figure}

\begin{figure*}
    \centering
    \includegraphics[width=0.48\textwidth]{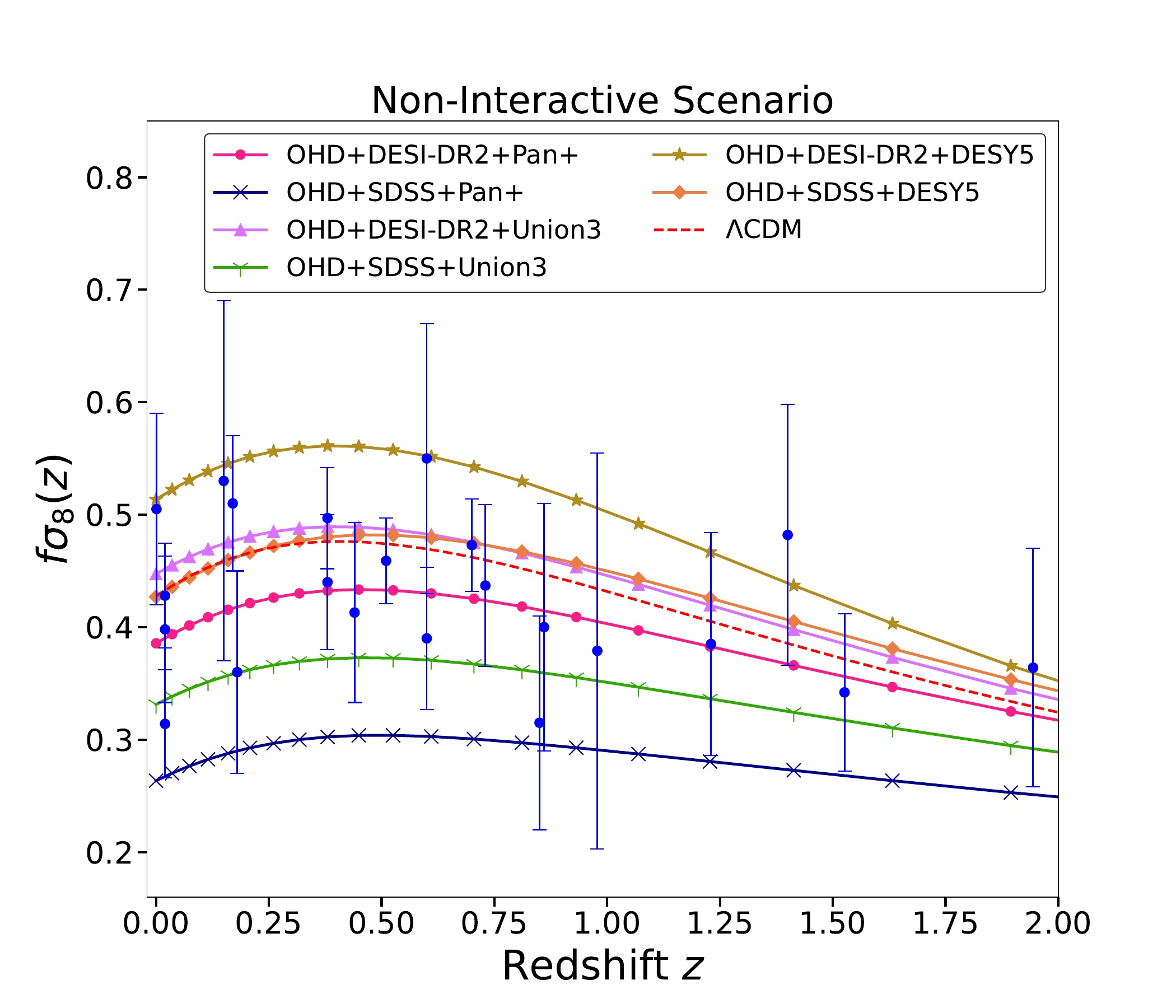}
    \includegraphics[width=0.48\textwidth]{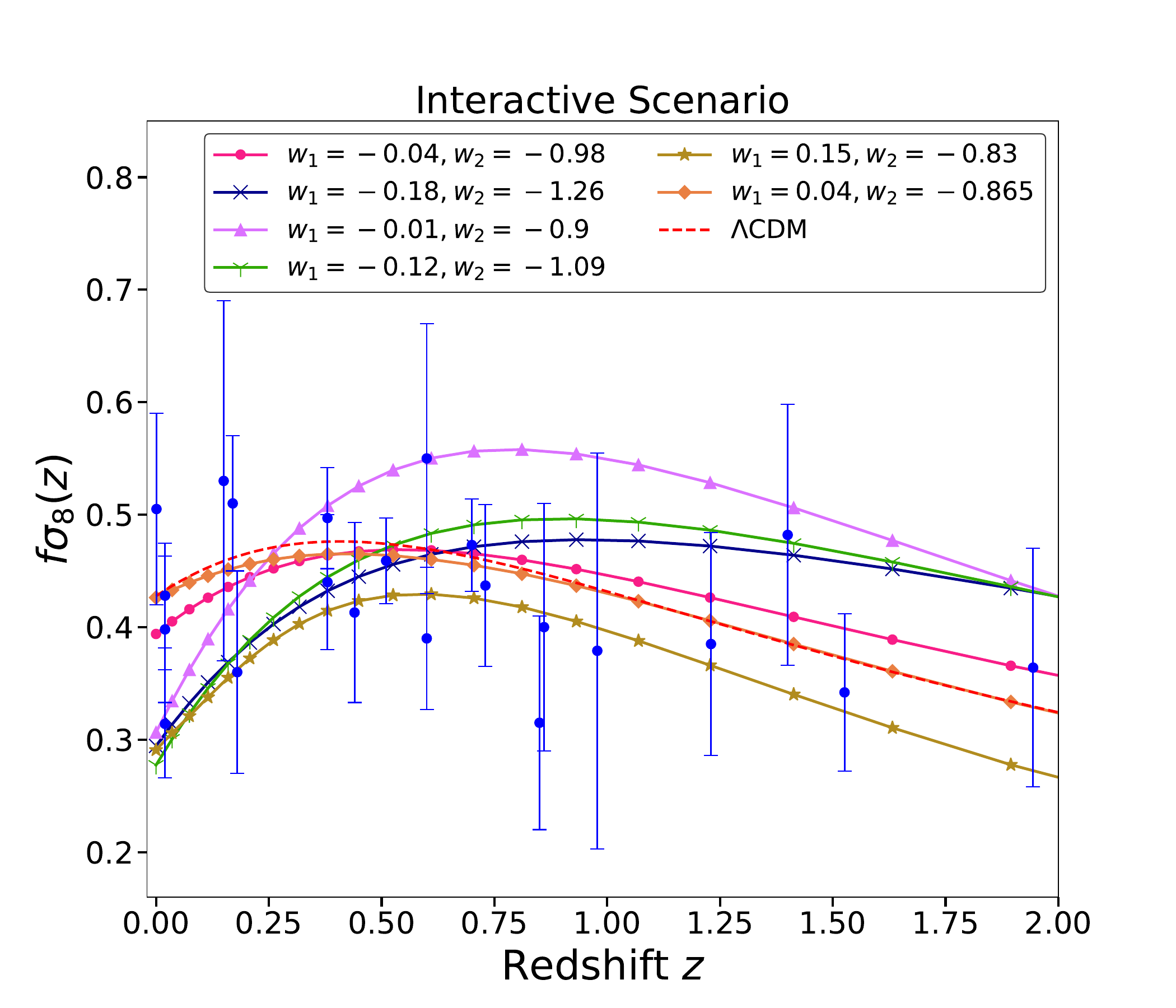}
    \caption{Evolution of growth rate parameter $f\sigma_8 (z)$ of our model and the concordance $\Lambda$CDM model under the non-interacting (\textbf{left}) and interacting scenarios (\textbf{right}), compared with the latest compilation of the $f\sigma_8$ measurements from various astronomical surveys, as summarized in Table~\ref{tab:fsigma8}.}
    \label{fig:fsigma8}
\end{figure*}

The constraints obtained at 68\% and 95\% CL for all data combinations are shown in Table~\ref{tab:Psi-mcmc}. In addition, Figs. \ref{fig:contour} and  \ref{fig:delta-alpha} present the graphical representation of two-dimensional (2D) contour plots of various free and derived parameters. Figure~\ref{fig:Eos-today} shows a whisker plot illustrating the 68\% CL constraints on the present-day value of the total equation of state parameter, $w_{\rm t, 0}$. Furthermore, in Fig.~\ref{fig:fsigma8}, we compare the predicted $f\sigma_8(z)$ evolution from both the proposed model and the $\Lambda$CDM model against recent observational measurements.

We begin by analyzing the constraints on the model parameters $\delta$ and $\alpha$. These two parameters are crucial, as deviations from their null values suggest the possibility of an alternative cosmological evolution beyond the standard $\Lambda$CDM-like evolution. Considering the OHD+SNIa datasets, the mean values of both $\delta$ and $\alpha$ show a deviation from the null, regardless of the choice of SNIa datasets. Notably for OHD+Pan$+$, there is a moderate evidence of $\delta \neq 0$ yielding $\delta = -0.35^{+0.11}_{-0.31}$ at 68\% CL--exceeding $1 \sigma$ level--compared to the results obtained using Union3 and DESY5. In the case of parameter $\alpha$, a mild yet notable deviation beyond $1\sigma$ level is observed in the case of OHD+DESY5, with $\alpha = 0.56^{+0.53}_{-0.24}$ at 68\% CL, as compared to the results from Pan$+$  and Union3.

The inclusion of the DESI-DR1 and DESI-DR2 dataset in the OHD+SNIa combination significantly enhances the constraints on the model parameters, as visualized in Fig.~\ref{fig:delta-alpha}. However, incorporating the more recent DESI-DR2 data alongside OHD+SNIa does not result in substantial changes in parameter constraints when compared to those obtained using DESI-DR1 alone in our model. Notably, both the OHD+DESI-DR1+DESY5 and OHD-DESI-DR2+DESY5 combinations yield strong evidence, with $\delta$ exceeding the $1\sigma$ level, and $\alpha$ showing a deviation  at $\sim 2\sigma$. In contrast, the combinations OHD+DESI-DR1+Pan$+$, OHD+DESI-DR2+Pan$+$, OHD+DESI-DR1+Union3, and OHD+DESI-DR2+Union3 indicate nonzero mean values for both $\delta$ and $\alpha$, but within $\sim 1\sigma$ range.

For $\delta$, a deviation exceeding the $1\sigma$ level is observed in the combinations OHD+SDSS+Pan$+$ and OHD+SDSS+Union3. In contrast, for the parameter $\alpha$, only the OHD+SDSS+DESY5 combination shows evidence beyond the $1\sigma$ level. However, a noticeable suppression in the mean values of both $\alpha$ and $\delta$ is observed across all OHD+SDSS+SNIa when compared to both OHD+DESI-DR1+SNIa and OHD+DESI-DR2+SNIa, regardless of the specific SNIa datset. This trend can be attributed to the positive correlation between $\alpha$ and $\delta$, as depicted in Fig. \ref{fig:delta-alpha}.

 Turning to the constraint on the Hubble parameter, we start with the OHD+SNIa combinations. At 68\% CL, the estimated values of $H_0$ are $H_0 = 67.02^{+0.85}_{-0.85}$ km/s/Mpc for OHD+Pan+,  $H_0 = 66.50 ^{+1.1}_{-1.1}$ km/s/Mpc for OHD+Union3 and $H_0 = 69.67^{+0.37}_{-0.37}$ km/s/Mpc for OHD+DESY5. 
The results from Pan+ and Union3 are in good agreement with Planck (assuming $\Lambda$ paradigm in the background)~\cite{Planck:2018nkj}, although they are accompanied by slightly larger uncertainties. In contrast, $H_0$ obtained from OHD+DESY5 deviates from the $\Lambda$CDM-based Planck~\cite{Planck:2018nkj} at several standard deviations. A similar trend in $H_0$ is observed when either DESI or SDSS is added to the OHD+SNIa combinations.

Focusing on the parameter $\Omega_0$, which plays a crucial role due to its correlation with other model parameters, we find relatively higher values for the OHD+Pan+ ($\Omega_0 = 0.458^{+0.170}_{-0.097}$ at 68\% CL) and OHD+Union3 ($\Omega_0 = 0.37^{+0.20}_{-0.16}$ at 68\% CL) combinations. In contrast, the value of $\Omega_0$ is noticeably lower in the OHD+DESY5 dataset, accompanied by reduced uncertainties. Although the inclusion of DESI or SDSS in the OHD+SNIa combinations leads to slight shifts in the mean values of $\Omega_0$, these variations are not statistically significant. 
Considering correlations among parameters, in the OHD+Union3 combination, the correlation between $\Omega_0$ and $H_0$ is not very clear. However, with the inclusion of DESI and SDSS in this combination, the $\Omega_0 - H_0$ contour plots tends to show a positive correlation. In all other combinations, positive correlations in the $\Omega_0 - H_0$ plane are quite evident, and the presence of DESI and SDSS strengthens these positive correlations. The relatively smaller contour area in the presence of DESI also highlights its stronger constraining power.

In the vanilla concordance model ($\Lambda$CDM), a negative correlation between $\Omega_0$ and $H_0$ is typically observed. However, our model reveals the opposite behavior. Several recent studies~\cite{Li:2024qus,Sabogal:2024yha,Chan-GyungPark:2024spk,Zheng:2024qzi} have also reported this reverse trend. Our model incorporates external degrees of freedom, which may introduce degeneracies in the parameter space. The observed reversal in the $\Omega_0 - H_0$ correlation may be indicative of such a degeneracy. To resolve this, stronger constraining power is required to break the degeneracy and more accurately determine the true nature of the correlation. Additionally, we find that $\Omega_0$ is negatively correlated with both model parameters $\delta$ and $\alpha$.

Furthermore, based on the estimated model parameters, the derived present-day value of the effective equation of state parameter, $w_{\rm t, 0}$, is illustrated in Fig. \ref{fig:Eos-today} as a 68\% CL whisker plot for all dataset combinations. At the current epoch, $w_{\rm t, 0}$ exhibits a significant deviation from the phantom divide, regardless of the dataset used, thereby indicating a quintessential nature of the effective EoS. This finding aligns with the theoretical predictions for the present behavior of the effective EoS, as discussed in Sec.~\ref{sec-2}.

Building upon Sec.~\ref{subsec-perturbed}, we present the redshift evolution of the growth rate parameter $f\sigma_8(z)$ for both our model and the $\Lambda$CDM model, compared against recent observational measurements of $f\sigma_8$ (see Table~\ref{tab:fsigma8}). These comparisons are shown in Fig.~\ref{fig:fsigma8}, where the left panel corresponds to the noninteractive scenario and the right panel depicts the interactive scenario. For our model, we use parameter priors derived from the combined observational constraints outlined in Sec.~\ref{sec-cons}, while for the $\Lambda$CDM model, we adopt the Planck 2018 prior as detailed in Ref.~\cite{Planck:2018vyg}. In the noninteractive scenario (left panel of Fig.~\ref{fig:fsigma8}), the predictions of our model, constrained by the OHD+DESI-DR2+Union3 and OHD+SDSS+DESY5 datasets, closely follow the $\Lambda$CDM curve. However, slight deviations from the observational $f\sigma_8$ data are observed within a narrow redshift range. For other dataset combinations, our model shows improved agreement with the data, capturing a broader range of the \( f\sigma_8 \) measurements. In the interactive scenario (right panel of Fig.~\ref{fig:fsigma8}), we investigate different theoretical choices for the EoS parameters, $w_1$ and  $w_2$, corresponding to two distinct fluids. Each parameter choice is depicted for specific combinations of the considered datasets. For instance, with $w_1 = 0.04$  and $w_2 = -0.865$, constrained by the OHD+SDSS+DESY5 dataset, our model provides a good fit to the \(\Lambda\)CDM model. However, for alternative combinations of $w_1$ and $w_2$, associated with different observational datasets, our model accommodates data points across various redshift bins. This clearly highlights the flexibility afforded by the choice of $w_1$ and $w_2$ in the presence of interaction between the two fluids, as well as the influence of different observational datasets, enabling a significantly improved fit to the recent $f\sigma_8$ measurements compared to the standard $\Lambda$CDM model.
\begin{table}
\centering
 \resizebox{0.8\columnwidth}{!}{
\small
\renewcommand{\arraystretch}{0.7}
\begin{tabular}{l c c c} 
\hline 
\multicolumn{4}{c}{\vspace{0.01mm}} \\
\textbf{Dataset} & \textbf{$z$} & $\mathbf{f\sigma_8(z)}$ & \textbf{Ref.} \\
\multicolumn{4}{c}{\vspace{0.01cm}} \\
\hline\hline 
\multicolumn{4}{c}{\vspace{0.01mm}} \\
~2MTF~ & $0.001$ & $0.505 \pm 0.085$ & \cite{Howlett:2017asq} \\
~6dFGS+SNIa~ & $0.020$ & $0.4280 \pm 0.0465$ & \cite{Huterer:2016uyq}\\
~IRAS+SNIa~ & $0.020$ & $0.398 \pm 0.065$ & \cite{Hudson:2012gt, Turnbull:2011ty}\\
~2MASS~ & $0.020$ & $0.314 \pm 0.048$ & \cite{Hudson:2012gt, Davis:2010sw}\\
~2dFGRS~ & $0.170$ & $0.510 \pm 0.060$ & \cite{Song:2008qt}\\
~GAMA~ & $0.180$ & $0.360 \pm 0.090$ & \cite{Blake:2013nif}\\
~GAMA~ & $0.380$ & $0.440 \pm 0.060$ & \cite{Blake:2013nif}\\
~SDSS-IV (MGS)~ & $0.150$ & $0.530 \pm 0.160$ & \cite{eBOSS:2020yzd} \\
~SDSS-IV (BOSS Galaxy)~ & $0.380$ & $0.497 \pm 0.045$ & \cite{eBOSS:2020yzd}\\
~SDSS-IV (BOSS Galaxy)~ & $0.510$ & $0.459 \pm 0.038$ & \cite{eBOSS:2020yzd}\\
~SDSS-IV (eBOSS LRG)~ & $0.700$ & $0.473 \pm 0.041$ & \cite{eBOSS:2020yzd}\\
~SDSS-IV (eBOSS ELG)~ & $0.850$ & $0.315 \pm 0.095$ & \cite{eBOSS:2020yzd}\\
~WiggleZ~ & $0.44$ & $0.413 \pm 0.080$ & \cite{Blake:2012pj} \\
~WiggleZ~ & $0.600$ & $0.390 \pm 0.063$ & \cite{Blake:2012pj}\\
~WiggleZ~ & $0.730$ & $0.437 \pm 0.072$ & \cite{Blake:2012pj}\\
~Vipers PDR-2~ & $0.600$ & $0.550 \pm 0.120$ & \cite{Pezzotta:2016gbo} \\
~Vipers PDR-2~ & $0.860$ & $0.400 \pm 0.110$ & \cite{Pezzotta:2016gbo} \\
~FastSound~ & $1.400$ & $0.482 \pm 0.116$ & \cite{Okumura:2015lvp} \\
~SDSS-IV (eBOSS Quasar)~ & $0.978$ & $0.379 \pm 0.176$ & \cite{eBOSS:2018yfg}\\
~SDSS-IV (eBOSS Quasar)~ & $1.230$ & $0.385 \pm 0.099$ & \cite{eBOSS:2018yfg}\\
~SDSS-IV (eBOSS Quasar)~ & $1.526$ & $0.342 \pm 0.070$ & \cite{eBOSS:2018yfg}\\
~SDSS-IV (eBOSS Quasar)~ & $1.944$ & $0.364 \pm 0.106$ & \cite{eBOSS:2018yfg}\\ 
\hline

\end{tabular}}
\caption{$f\sigma_8(z)$ measurements obtained from different observational surveys, which have been used in this work.}
 \label{tab:fsigma8}
\end{table}

\subsection{Model analysis through the lens of thermodynamics}
\label{sec-thermo}

In this section, we explore the thermodynamic properties of the proposed cosmological model, with an emphasis on the generalized second law of thermodynamics \cite{Radzikowski:1988kv,
Davies:1988dk,MohseniSadjadi:2007zq,
Horvat:2007jr,Setare:2008bb,Karami:2010zz}. The empirical basis of the second law of thermodynamics is based on the principle that macroscopic systems spontaneously evolve toward thermodynamic equilibrium. According to this principle, the entropy ($S$) of an isolated system never decreases, implying, $dS/dt\geq 0$, and should be concave, $d^2S/dt^2 < 0$ as the system approaches thermodynamic equilibrium. 

To study the thermodynamic behavior for cosmological scenarios, we model the Universe as a thermodynamic system bounded by the apparent horizon. Within the FLRW framework, the total entropy, comprising the entropy of the proposed field ($S_{f}$) and the entropy of the apparent horizon ($S_{h}$), must satisfy the condition $d{S}_{\rm total}/dt = d{S}_{f}/dt + d{S}_{h}/dt \geq 0 $. Furthermore, at later times, $d^2{S}_{f}/dt^2 + d^2{S}_{h}/dt^2 < 0 $ should hold. As outlined in Ref.~\cite{Jamil:2009eb}, the cosmic time derivatives of the entropy of the apparent horizon ($d{S}_{h}/dt$) and the proposed fluid ($d{S}_{f}/dt$) are expressed as 
\begin{eqnarray}
 \frac{d{S}_{h}}{dt} &=& 16 \pi^{2} r_{A}\frac{d{r}_{A}}{dt}, \label{eq:S_dot_h}\\
 \frac{d{S}_{f}}{dt} &=& \frac{1}{T}\left(P_{f}\cdot 4\pi r_{A}^{2} \frac{d{r}_{A}}{dt} + \frac{d{E}_{f}}{dt} \right). \label{eq:S_dot_f}
\end{eqnarray}
Here, the apparent horizon radius $r_A$ is defined as $r_A = 1/H$ for a flat Universe. The temperature of the Universe $T$ is given by $1/(2 \pi r_{A})$. The pressure of the dark energy fluid is denoted by $P_f$, while the energy of the field $f$ is expressed as $E_f = \frac{4\pi}{3}r_{A}^{3}\rho_{f}$ where $\rho_f$ is the energy density. Using Eqs.~\eqref{eq:S_dot_h} and~\eqref{eq:S_dot_f}, the cosmic time derivative of the total entropy is written as

\begin{equation}\label{eq:S_dot_total}
\frac{d{S}_{\rm total}(z)}{dt} = \frac{16 \pi^{2}}{H^{5}} \left(\frac{dH}{dt}\right)^2,  
\end{equation}
while the second cosmic time derivative of the total entropy [$d^2{S}_{\rm total}(z)/dt^2$] becomes

\begin{equation}\label{eq:S_ddot_total}
    \frac{d^2{S}_{\rm total}(z)}{dt^2} = \frac{16 \pi^2}{H^{5}} \bigg(  2 \frac{dH}{dt}\frac{d^2H}{dt^2} - \frac{5}{H} \left(\frac{dH}{dt}\right)^3\bigg).
\end{equation}

\begin{figure*}
    \centering
    \includegraphics[width=0.48\textwidth]{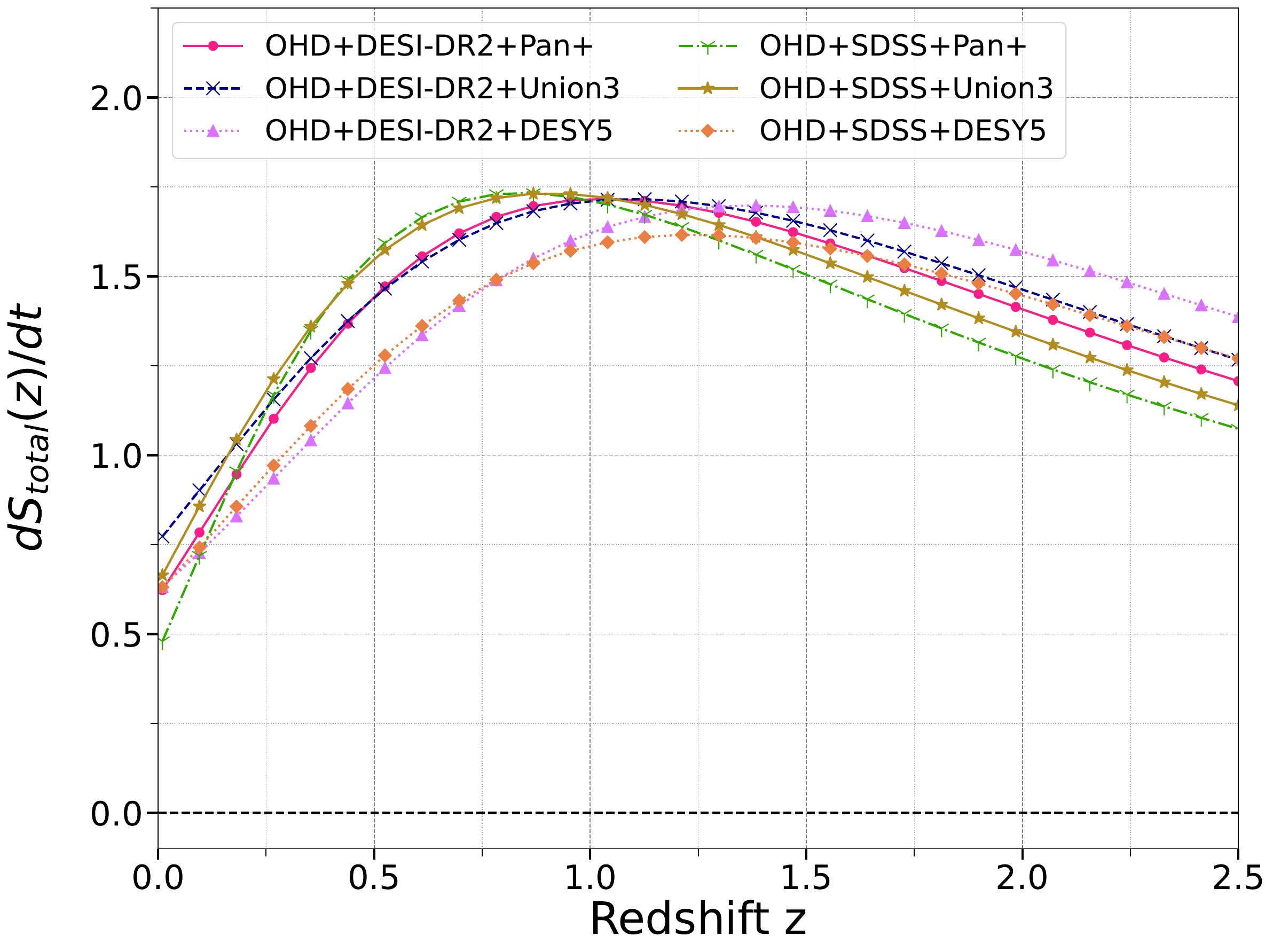}
    \includegraphics[width=0.48\textwidth]{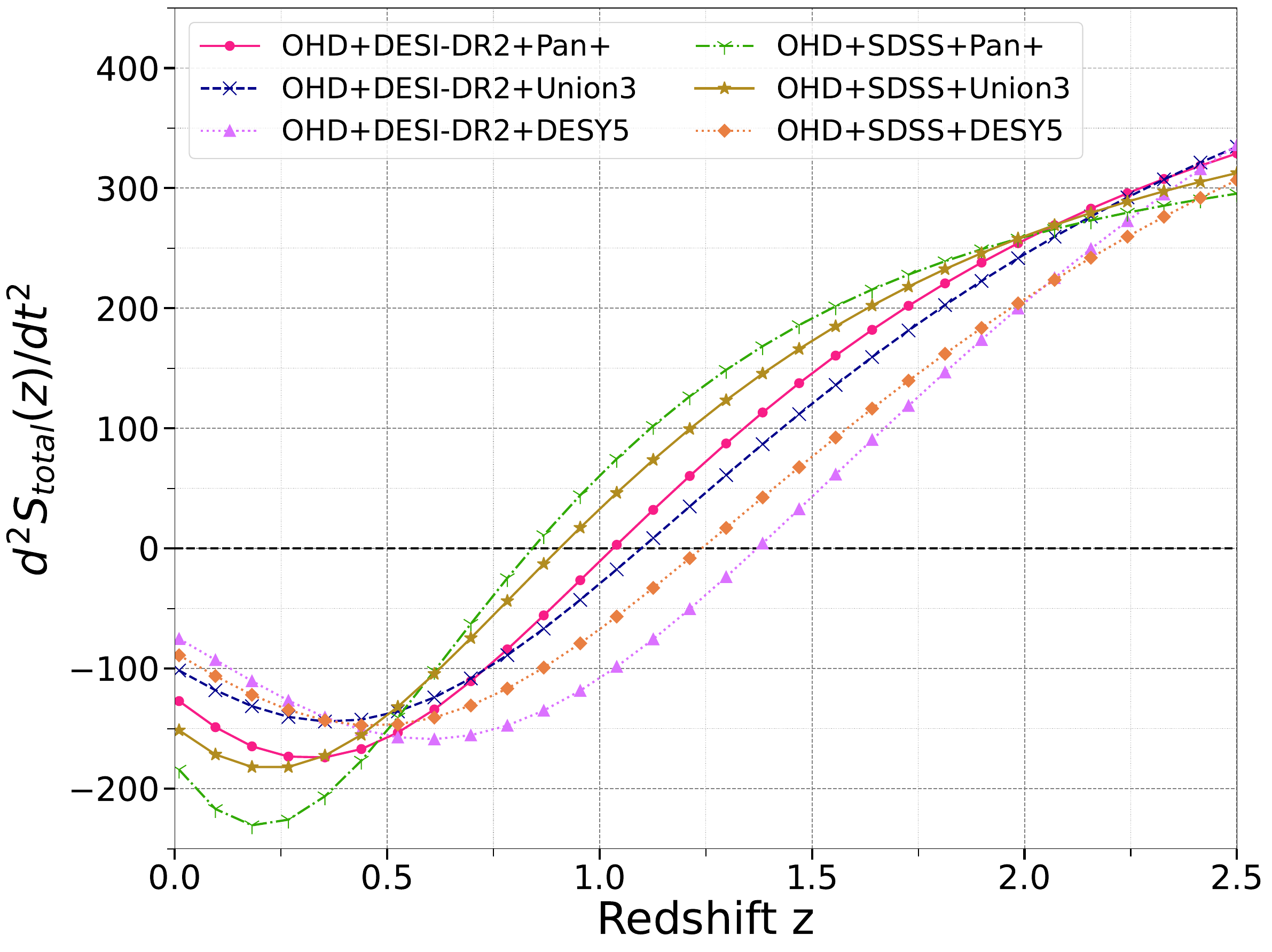}
    \caption{ Redshift evolution of $dS_{\rm total}(z)/dt$ (\textbf{left}) and $d^2S_{\rm total}(z)/dt^2$ (\textbf{right}) for our model, considering all dataset combinations. The black dashed lines indicate the thermodynamic thresholds: $dS_{\rm total}(z)/dt \geq 0$ (\textbf{left}) and $d^2S_{\rm total}(z)/dt^2 < 0$ (\textbf{right}), consistent with second law of thermodynamics at the present epoch. }
    \label{fig:evolve}
\end{figure*}
\begin{figure*}
    \centering    \includegraphics[width=1.0\textwidth]{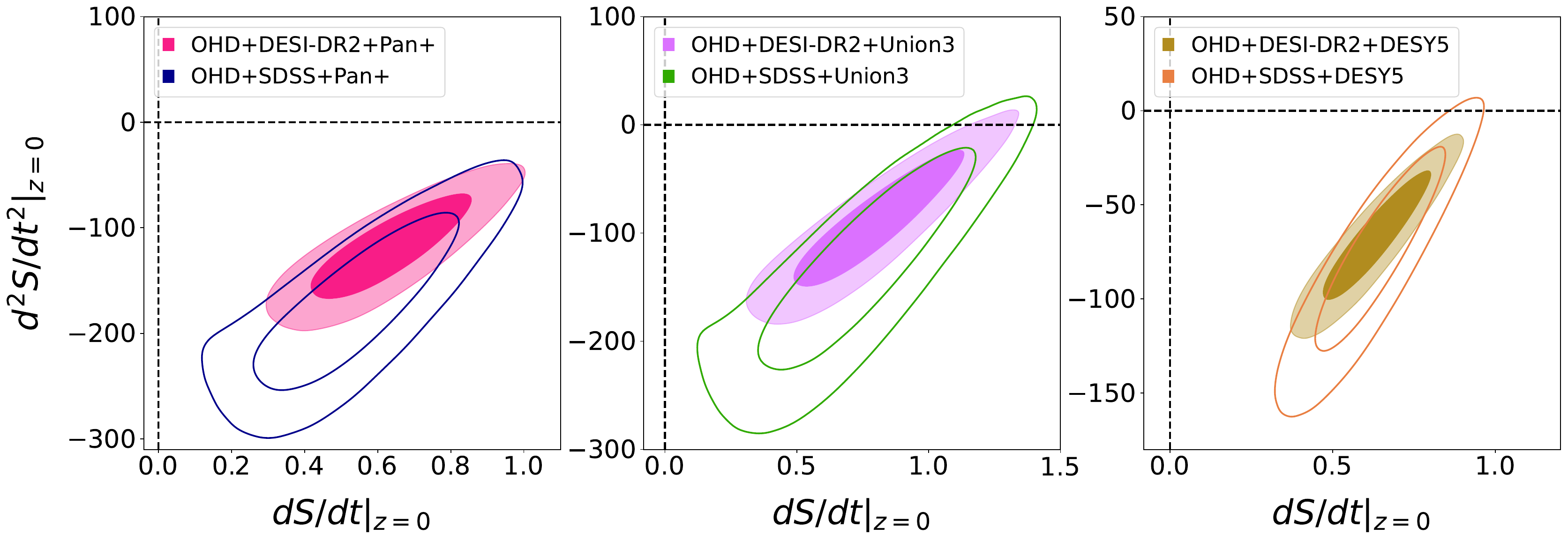}
    \caption{The figure illustrates the two-dimensional marginalized contour in the ($d S/dt|_{z=0}~-~d^2 S /dt^2|_{z=0}$) for different observational datasets analyzed in this article.  The vertical line at $d S/dt|_{z=0}$ and the horizontal line at $d^2 S /dt^2|_{z=0}$ serve as thresholds delineating the conditions for adherence to the second law of thermodynamics, namely $d S/dt|_{z=0} \geq 0$ and $d^2 S /dt^2|_{z=0} < 0$.}
    \label{fig:2d_contour_Entropy}
\end{figure*}

In the subsequent part of this section, we analyze the evolution of the first and second cosmic time derivatives of the thermodynamic observable entropy, $d S/dt$ and $d^2 S /dt^2$, using constraints from various observational datasets:  OHD+DESI-DR2+Pan$+$, OHD+DESI-DR2+Union3, OHD+DESI-DR2+DESY5, OHD+SDSS+Pan$+$, OHD+SDSS+Union3, and OHD+SDSS+DESY5 as illustrated in Fig.~\ref{fig:evolve}.~\footnote{As discussed in Sec.~\ref{sec-cons}, DESI-DR1 and DESI-DR2 yield comparable parameter constraints; we present the evolution of $dS/dt$ and $d^2 S/dt^2$ using only DESI-DR2.} The resulting behavior validates the second law of thermodynamics within the context of our model. In all dataset combinations, $d S(z)/ dt$ remains positive ($d S(z)/dt > 0$) and reaches a maximum in the region $0.5 \leq z \leq 2.0$. The curves for OHD+SDSS+DESY5 and OHD+DESI-DR2+DESY5 reach their maximum values at relatively early times, around $z \thicksim 1.5$.

The evolution of  $d^2 S(z)/ dt^2$ validates the second law of thermodynamics, with $d^2 S(z)/ dt^2 < 0$ at late times.  As illustrated in Fig.~\ref{fig:evolve}, a transition from positive to negative values occurs within the redshift range $0.9 \leq z \leq 1.5$, indicating the onset of thermodynamic equilibrium. The most pronounced negative values appear between $z = 0.25$ and $z = 0.75$. At the present epoch ($z = 0$), $d^2S/dt^2$ consistently remains negative, ranging from $-150$ to $-50$ across all dataset combinations. However, as seen in previous cases, OHD+DESI-DR2+DESY5 and OHD+SDSS+DESY5 combinations reach their minima slightly earlier, around $z \thicksim 0.7$.

The behavior of $d^2 S(z)/dt^2$ reveals a transition from positive values at higher redshift to negative values at lower redshift, reflecting the Universe's progression from a matter-dominated era [characterized by $d^2 S(z)/dt^2 > 0$] to a DE dominated era [where $d^2 S(z)/dt^2 < 0$]. Furthermore, the gradual stabilization of $d^2 S(z)/dt^2$ to a specific negative value at lower redshift suggests that the model approaches thermodynamic equilibrium in the late-time regime, consistent with the predictions of the second law of thermodynamics.

 Further, we constrain the $d S/dt|_{z=0}$ and $d^2 S /dt^2|_{z=0}$ as illustrated in Fig.~\ref{fig:2d_contour_Entropy}. Each plot illustrates a comparison of the 2D contours in the $d S/dt|_{z=0} ~ - ~ d^2 S /dt^2|_{z=0}$ plane between the combinations OHD+DESI-DR2+SNIa and OHD+SDSS+SNIa. Starting with the constraints derived from OHD+SDSS+Pan+, the contour resides within the region consistent with the second law of thermodynamics. The inclusion DESI-DR2 retains this behavior, remaining within the defined thresholds while offering significantly tighter constraints. In contrast, the combination of OHD+SDSS+Union3 and OHD+SDSS+DESY5 show a deviation from the second law of thermodynamics at $\sim 1\sigma$ CL. Notably, when DESI-DR2 is included with DESY5, the deviation becomes more pronounced, reaching a significance level $> 2\sigma$, accompanied by tighter constraints. However, the combination OHD+DESI-DR2+Union3 yields tighter constraints without a notable change in the level of deviation compared to the SDSS counterpart. 

These variations in constraining power among the datasets can be attributed to the properties of the respective SNIa samples. Pan+ spans a broad redshift range, offering stronger constraints. In contrast, Union3 includes only 22 binned data points, and DESY5 is confined to a limited redshift region, both of which restrict their effectiveness in providing tighter constraints on $d S/dt|_{z=0}$ and $d^2 S/ dt^2|_{z=0}$.

\subsection{Statistical insights on model selection}
\label{sec-model-comparison}

\begin{table*}[ht]
\centering
\renewcommand{\arraystretch}{1.2}
\resizebox{1.0\textwidth}{!}{\huge \begin{tabular}{l @{\hspace{0.5 cm}} l @{\hspace{0.5 cm}} c @{\hspace{0.5 cm}} c @{\hspace{0.5 cm}} c @{\hspace{0.5 cm}} c @{\hspace{0.5 cm}} c @{\hspace{0.5 cm}} c @{\hspace{0.5 cm}} c @{\hspace{0.5 cm}} c  @{\hspace{0.5 cm}} c  @{\hspace{0.5 cm}} c}

\toprule
\hline 
\textbf{Dataset} & \textbf{Model} & \textbf{AIC} & \boldmath{$\Delta$AIC} & \textbf{BIC} & \boldmath{$\Delta$BIC} & \textbf{$\chi^2$} & \boldmath{$\Delta \chi^2$} & \boldmath{DIC}  & \boldmath{$\Delta $DIC}  & \boldmath{$\ln{Z}$} & \boldmath{$\Delta \ln{Z}$} \\

\hline\hline

\multirow{2}{*} {OHD+Pan+} & our model & $1444.55$ & $-7.96$ & $1471.60$ & $2.86$ &  $1434.55$ & $-11.96$ & $1438.79$ &  $-14.12 $ & $-733.850~(0.119)$ & $-1.441~(0.016) $ \\
 
 & $\Lambda$CDM & $1452.51$ & $-$ &  $1468.74$ & $-$ & $1446.51$ & $-$ & $1452.51$ & $-$ &$-735.291~(0.103)$ & $-$\\
 
\multicolumn{12}{ c }{\hdashrule{\dimexpr\textwidth-2\tabcolsep\relax}{0.5pt}{3mm 2mm}}\\

\multirow{2}{*}{OHD+DESI-DR1+Pan+} & our model & $1466.03$ & $-3.07$ & $1493.12$  & $7.77$ & $1456.03$ & $-7.07$ & $1463.05$ & $-6.08$ & $-745.968~(0.125)$ & $1.854~(0.020) $\\
 
 & $\Lambda$CDM & $1469.10$ & $-$ & $1485.35$ & $-$ & $1463.10$ & $-$ & $1469.13$ & $-$ & $-744.114~(0.105)$ & $-$\\

\multicolumn{12}{ c }{\hdashrule{\dimexpr\textwidth-2\tabcolsep\relax}{0.5pt}{3mm 2mm}}\\

\multirow{2}{*} {OHD+DESI-DR2+Pan+} & our model & $1464.00$ & $-3.27$ & $1491.09$  & $7.57$ & $1454.00$ & $-7.27$ & $1462.30$ & $-4.96$ & $-746.240~(0.130)$ & $2.016~(0.020)$\\
 
 & $\Lambda$CDM & $1467.27$ & $-$ & $1483.52$ & $-$ & $1461.27$ & $-$ & $1467.26$ & $-$&$-744.224~(0.110)$ & $-$\\

\multicolumn{12}{ c }{\hdashrule{\dimexpr\textwidth-2\tabcolsep\relax}{0.5pt}{3mm 2mm}}\\

\multirow{2}{*} {OHD+SDSS+Pan+} & our model & $1452.56$ & $-10.55$ & $1479.65$  & $0.29$ & $1442.56$ & $-14.55$ & $ 1448.59$  &  $-14.51$ &  $-738.799~(0.123)$ & $-2.193~(0.018)$\\
 
 & $\Lambda$CDM & $1463.11$ & $-$ & $1479.36$ & $-$ & $1457.11$ & $-$ & $1463.10$  & $-$ & $-740.992~(0.105)$ & $-$\\
 
\hline

\multirow{2}{*}{OHD+Union3} & our model & $66.29$ & $-7.76$  & $76.01$ & $-2.90$ &  $58.29$  & $-11.76 $ & $ 59.11$ & $-14.94$ &  $-40.647~ (0.098)$ & $-1.703~ (0.019)$ \\
 
 & $\Lambda$CDM & $74.05$  & $-$ & $78.91$ & $-$ & $70.05$ & $-$ & $74.05$ & $-$ & $-42.350~(0.079)$ & $-$ \\

\multicolumn{12}{ c }{\hdashrule{\dimexpr\textwidth-2\tabcolsep\relax}{0.5pt}{3mm 2mm}}\\

\multirow{2}{*} {OHD+DESI-DR1+Union3} & our model & $87.07$ & $-3.20$ & $97.33$ & $1.93 $ & $79.07$ & $-7.20$ & $80.36$ & $-9.91 $ & $-52.689 ~(0.106)$ & $1.565~ (0.022)$\\
 
 & $\Lambda$CDM &$90.27$  & $-$ &  $95.40$& $-$ & $86.27$ & $-$ & $90.27$ & $-$ &$-51.124~ (0.084)$ & $-$\\

\multicolumn{12}{ c }{\hdashrule{\dimexpr\textwidth-2\tabcolsep\relax}{0.5pt}{3mm 2mm}}\\

\multirow{2}{*} {OHD+DESI-DR2+Union3} & our model & $84.77$ & $-3.86$ & $95.07$  & $1.29$ & $76.77$ & $-7.86$ & $77.71$ & $-10.92$  &$-52.610~(0.111)$ & $1.657~(0.024) $\\
 
 & $\Lambda$CDM & $88.63$ & $-$ & $93.78$ & $-$ & $84.63$ & $-$ &  $88.63$ & $-$ & $-50.953~(0.087)$ & $-$\\

\multicolumn{12}{ c }{\hdashrule{\dimexpr\textwidth-2\tabcolsep\relax}{0.5pt}{3mm 2mm}}\\

\multirow{2}{*} {OHD+SDSS+Union3} & our model & $74.56$ & $-10.16$ & $84.90$ & $-5.03$ & $66.56$ & $-14.16$ &  $69.94$ & $-14.78 $  & $-45.706~ (0.084)$ & $-2.438~ (0.017)$\\
 
 & $\Lambda$CDM & $84.72$ & $-$ & $89.89$ & $-$ & $80.72$ & $-$ & $84.72$ & $-$ & $-48.144~ (0.067)$ & $-$ \\
 
\hline

\multirow{2}{*}{OHD+DESY5}& our model & $1700.00$ & $-12.89$ & $1722.18$  & $-1.80$ & $1692.00$ & $-16.89$ & $1678.48$ & $-34.41$ & $-860.349~ (0.091)$ & $-3.036~ (0.018)$\\
 
 & $\Lambda$CDM & $1712.89$  & $-$ & $1723.98$ & $-$ & $1708.89$ & $-$ & $1712.89$ &$-$ &$-863.385~ (0.073)$ & $-$\\
 
\multicolumn{12}{ c }{\hdashrule{\dimexpr\textwidth-2\tabcolsep\relax}{0.5pt}{3mm 2mm}}\\

\multirow{2}{*} {OHD+DESI-DR1+DESY5} & our model & $1721.36$ & $-6.98$ & $1743.56$ & $4.11$ & $1713.36$ & $-10.98$ & $1702.83$ & $-25.53$ & $-872.395~ (0.117)$ & $0.863~ (0.042)$\\

 & $\Lambda$CDM & $1728.34$ & $-$ & $1739.45$ & $-$ & $1724.34$ & $-$ & $1728.36$ & $-$ & $-871.532~ (0.075)$ & $-$\\
 
\multicolumn{12}{ c }{\hdashrule{\dimexpr\textwidth-2\tabcolsep\relax}{0.5pt}{3mm 2mm}}\\

\multirow{2}{*} {OHD+DESI-DR2+DESY5} & our model & $1731.12$ & $-2.59$ & $1753.33$  & $8.52$ & $1723.12$ & $-6.59$ & $1723.44$ & $-10.26$ & $-877.430~(0.118)$ & $2.784~(0.024) $\\

 & $\Lambda$CDM & $1733.71$ & $-$ & $1744.81$ & $-$ & $1729.71$ & $-$ & $1733.70$ & $-$ & $-874.646~(0.094)$ & $-$\\

\multicolumn{12}{ c }{\hdashrule{\dimexpr\textwidth-2\tabcolsep\relax}{0.5pt}{3mm 2mm}}\\

 \multirow{2}{*}{OHD+SDSS+DESY5} & our model & $1722.34$ & $-6.00$ & $1744.55$ & $5.1$ & $1714.34$ & $-10.00$ & $1712.56$ & $-11.46$ & $-871.627~ (0.092) $ & $-1.575~ (0.018)$\\
 
 & $\Lambda$CDM & $1728.34$ & $-$ & $1739.45$ & $-$ & $1724.34$ & $-$ & $1724.02$ & $-$ & $-873.202~(0.074)$ & $-$ \\

\hline\hline
\bottomrule
\end{tabular}}
\caption{Statistical comparison of our model and $\Lambda$CDM using the minimum $\Delta \chi^2_{\text{min}}$, information criteria $\Delta$XIC (where $X = \text{A}, \text{B}, \text{D}$), and relative log-Bayesian evidence $\Delta \ln \mathcal{Z}$ across twelve dataset combinations. Negative values of $\Delta \chi^2_{\text{min}}$, $\Delta$XIC, and $\Delta \ln \mathcal{Z}$ indicate a preference for our model over the $\Lambda$CDM reference model.}
\label{tab:model_comparison}
\end{table*}
\begin{figure*}
    \centering
    \includegraphics[width=1.03\textwidth]{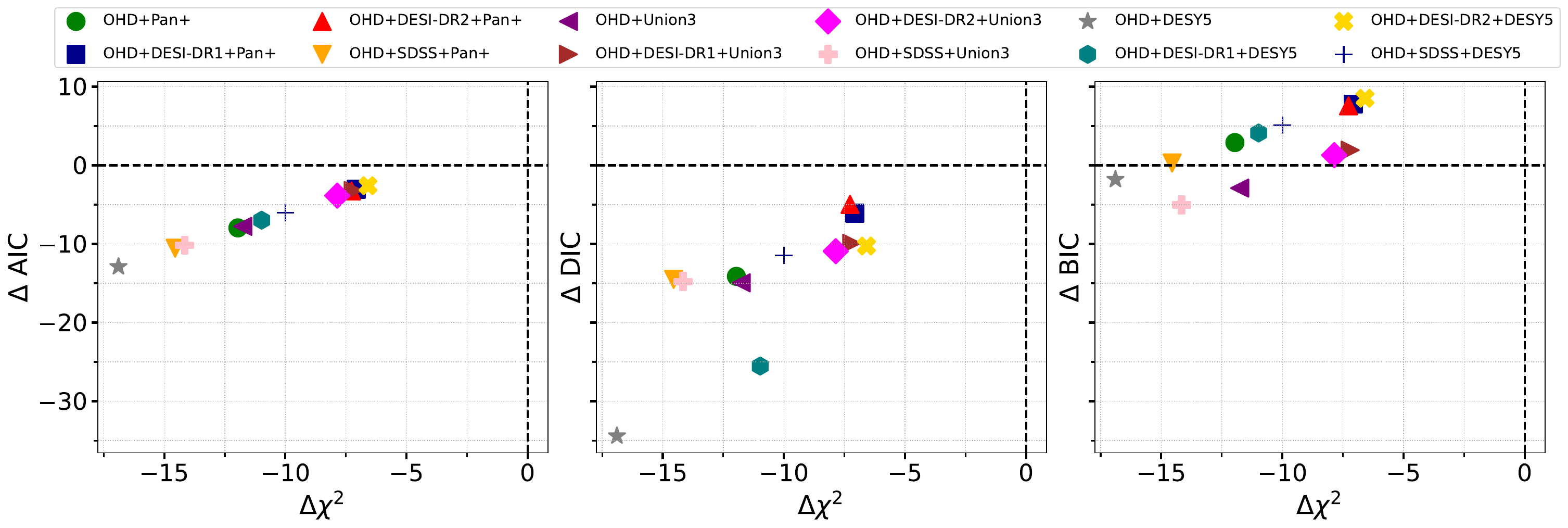}
    \caption{An overview of model selection results across datasets, quantified by the differences in the minimum chi-squared, $\Delta \chi^2_{\text{min}}$, and the information criteria $\Delta \text{AIC}$, $\Delta \text{DIC}$, and $\Delta \text{BIC}$ (\textbf{left}, \textbf{middle}, and \textbf{right} panels, respectively), for our model. In each case, the dashed line indicates the demarcation threshold, where negative values signify that our model is statistically preferred over the $\Lambda$CDM reference model.}
    \label{fig:Information_Criterion}
\end{figure*}
\begin{figure}
    \centering
    \includegraphics[width=0.48\textwidth]{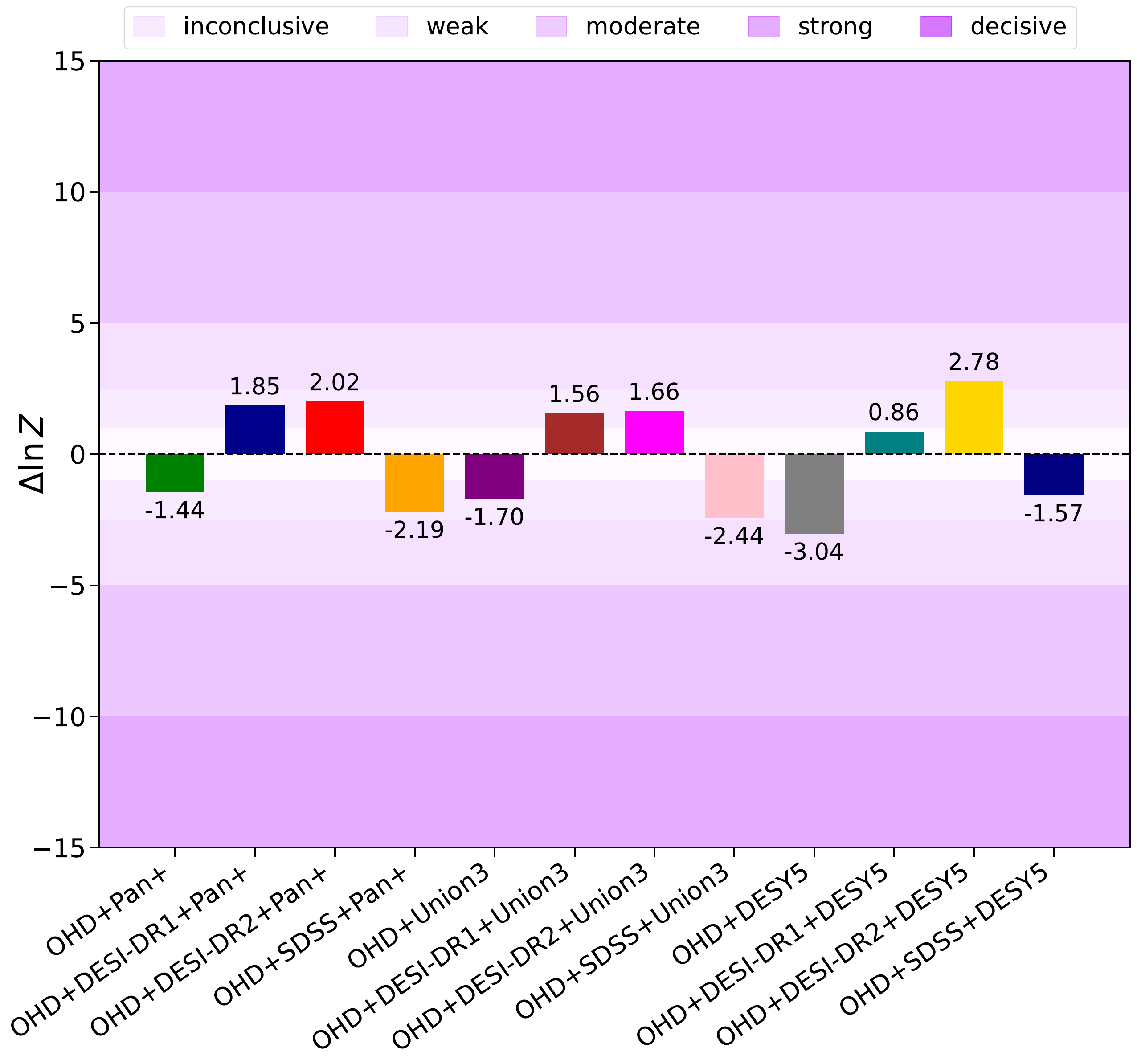}
    \caption{The relative log-Bayesian evidence in the form of $\Delta\ln Z$ draws the conclusion regarding model comparison for all the dataset combinations. Negative values of $\Delta \ln \mathcal{Z}$ indicate the preference for our model over the $\Lambda$CDM reference model. }
    \label{fig:Bayesian}
\end{figure}
\begin{figure*}
    \centering
     \includegraphics[width=1.06\textwidth]{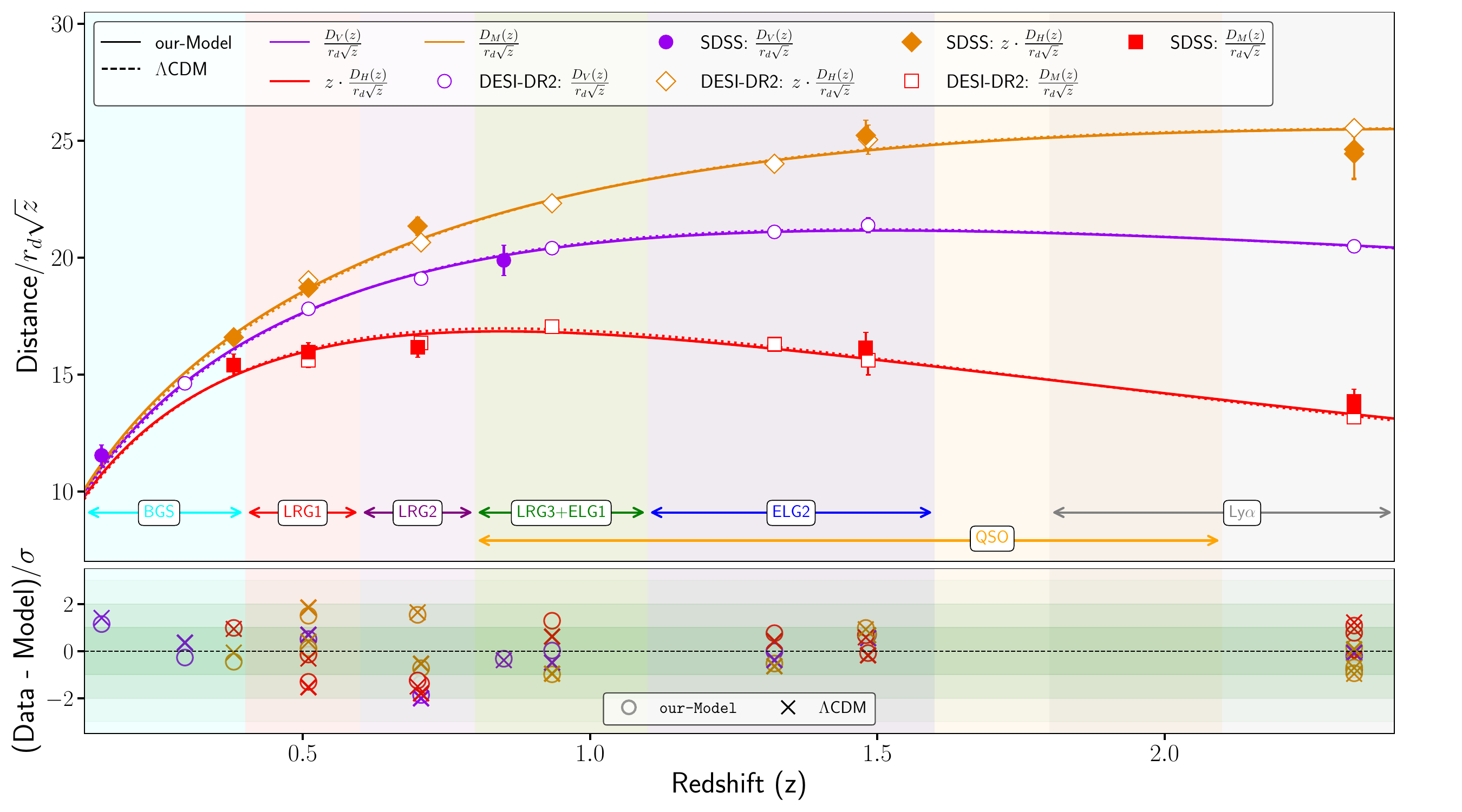}
    \caption{\textit{\textbf{Upper}}: rescaled distance-redshift relations predicted by our model (\textit{solid lines}) and the $\Lambda$CDM model (\textit{dashed lines}), derived using the combined constraints from OHD+DESI-DR2+Pan+. The curves correspond to the three rescaled BAO distance measures, color coded as indicated in the legend. Error bars denote the $\pm 1\sigma$ observational uncertainties. `\textit{empty}' markers represent DESI-DR2 measurements, while `\textit{filled}' marker correspond to data from SDSS. \textit{\textbf{Lower}}: normalized residuals between the theoretical predictions and BAO observations. `\textbf{o}' markers denote the deviation for our model, while `\textbf{x}' markers represent those for the $\Lambda$CDM model.}
    \label{fig:Distance}
\end{figure*}

With our key statistical and cosmological findings in place, we proceed to assess level of (dis)agreement between our model and the observational data. To perform the statistical comparison, we adopt the Akaike information criterion (AIC)~\cite{1100705, Trotta:2008qt}, the Bayesian information criterion (BIC)~\cite{10.1214/aos/1176344136}, the deviance information criterion (DIC)~\cite{Spiegelhalter:2002yvw, Liddle:2007fy, Kunz:2006mc}, and log-Bayesian evidence, along with $\chi^2_{min} = -2 \ln \mathcal{L}_{max}$, where $\mathcal{L}_{max}$ is the maximum likelihood, as illustrated in Table~\ref{tab:model_comparison}. More specifically, we are interested in evaluating relative best fit ($\Delta \chi^2_{min} = \chi^2_{min, \text{model}} - \chi^2_{min, \Lambda \text{CDM}}$), relative AIC ($\Delta \text{AIC} = \text{AIC}_{\rm model} - \text{AIC}_{\Lambda \text{CDM}}$, where  $\text{AIC} \equiv \chi^2_{min} + 2k$, where $k$ is total number of free parameters), relative BIC ($\Delta \text{BIC} = \text{BIC}_{\text{model}} - \text{BIC}_{\Lambda \text{CDM}}$, where $\text{BIC} \equiv \chi^2_{min} +  k\ln N$, with $N$ being the number of data points), and the relative DIC [$\Delta \text{DIC} = \text{DIC}_{\text{model}} - \text{DIC}_{\Lambda \text{CDM}}$, where  $\text{DIC} \equiv 2\overline{D(\theta)} - D(\overline{\theta})$, and $D(\theta) = -2\ln \mathcal{L}(\theta) + C $, with the overbar denoting the mean parameter likelihood and $C$ being a data-driven constant that cancels out from any derived quantity]. 

The support for the model is interpreted based on thresholds in $\Delta \text{AIC}$, $\Delta \text{BIC}$ and $\Delta \text{DIC}$ relative to a reference model. A $\Delta \text{AIC} < 2$ (in particular $< 0$) indicates strong support for the model over the reference model, while values between $4$ and $7$ reflect considerably less support, and $\Delta \text{AIC} > 10$ indicates essentially no support for the model \cite{burnham2003model}. Similarly, $\Delta \text{BIC} < 2$ provides no distinguishable evidence between models; values between 2 and  6  indicate mild to positive evidence against the model; $6< \Delta \text{BIC} < 10 $ represents strong evidence; and values $ > 10$ suggest very strong evidence favoring rejection of the model in favor of the reference model \cite{Kass:1995loi}. The interpretation of $\Delta \text{DIC}$ generally aligns with that of $\Delta \text{AIC}$~\cite{Rezaei:2021qpq}.

To further assess the evidence of the model, we analyze the relative log-Bayesian evidence, defined as $\Delta \ln \mathcal{Z} \equiv \ln \mathcal{B}_{ij} =  \ln \mathcal{Z}_{\Lambda\text{CDM}} - \ln \mathcal{Z}_{\text{model}}$, where $\mathcal{B}_{ij}$  is the Bayes factor and  $\mathcal{Z}_{i}$, $\mathcal{Z}_{j}$ represents the Bayesian evidences for models $i$ and $j$, respectively. The Bayesian evidence values were computed using the nested sampling algorithm implemented in the \texttt{\small DYNESTY}\footnote{\href{https://github.com/joshspeagle/dynesty.git}{https://github.com/joshspeagle/dynesty.git}} package. In interpreting the results, we apply the revised Jeffreys' scale~\cite{Kass:1995loi, Trotta:2008qt}, where $|\Delta\ln{Z}| \lesssim 1$ indicates inconclusive evidence, $1 \lesssim |\Delta\ln{Z}| \lesssim 3$ corresponds to a moderate evidence, and $3 \lesssim |\Delta\ln{Z}| \lesssim 5$ suggests strong evidence. In all cases, the negative values of $\Delta \chi^2$, $\Delta \text{AIC}$, $\Delta \text{BIC}$, $\Delta \text{DIC}$ and $\Delta \ln \mathcal{Z}$  imply a statistical preference for the proposed model over the reference model.

As shown in Table~\ref{tab:model_comparison}, we compare our model with the reference standard model of $\Lambda$CDM. For the estimation of $\Delta \chi^2$, $\Delta \text{AIC}$ and $\Delta \text{DIC}$ using the OHD+SNIa combination, we find strong support in favor of our model , regardless of the specific SNIa sample considered. Furthermore, with the inclusion of both DESI-DR1 and DESI-DR2, the level of support for our model remains consistently strong compared to the  $\Lambda$CDM model. A similar trend in favor of our model is also observed when the dataset is extended to include OHD+SDSS+SNIa.

In contrast, the evaluation based on $\Delta \text{BIC}$ reveals a partial shift in preference toward the reference model for certain dataset combinations. A preference  for  our model is observed when considering the OHD+Union3 and OHD+DESY5 combinations. However, in the case of OHD+Pan+ combination, mild support for the $\Lambda$CDM model is favored. With the inclusion of both DESI-DR1 and DESI-DR2,  the level of support appears mixed. In particular a strong preference for $\Lambda$CDM is observed in the OHD+DESI-DR1+Pan+ and OHD+DESI-DR2+Pan+ combinations, while for OHD+DESI-DR1+Union3 and OHD+DESI-DR2+Union3, no conclusive evidence is obtained. Nevertheless, when both DESI data are combined with OHD+DESY5, a mild to strong preference for $\Lambda$CDM is observed. Additionally, the combination of OHD+SDSS+Union3 shows strong support for our model, while OHD+SDSS+Pan+ yields results that are statistically indistinguishable between the models. In contrast, OHD+SDSS+DESY5 exhibits a strong preference for the $\Lambda$CDM model. The discrepancies observed in the $\Delta \text{BIC}$ outcomes, compared to those from $\Delta \chi^2$, $\Delta \text{AIC}$, and $\Delta \text{DIC}$, are primarily due to the fact that, the BIC imposes a stronger penalty on models with a greater number of free parameters than the other information criteria~\cite{Trotta:2017wnx}. To further understand the variation across different information criterion, the relationships between $\Delta \text{AIC}$, $\Delta \text{DIC}$, and $\Delta \text{BIC}$ relative to $\Delta \chi^2$ for various dataset combinations are shown in Fig.~\ref{fig:Information_Criterion}.

While considering the relative log-Bayesian evidence, $\Delta \ln \mathcal{Z}$, a moderate preference for our model is observed in the cases of the OHD+Pan+ and OHD+Union3 combinations, whereas strong evidence in favor of our model is found for OHD+DESY5. However, when both DESI datasets are combined with OHD+SNIa, the preference shifts moderately in favor of $\Lambda$CDM, except in the case of the OHD+DESI-DR1+DESY5 combination, where the evidence is inconclusive. A moderate support for our model is also observed when using the combination of OHD+SDSS+SNIa, with results similar to those obtained with OHD+SNIa alone. For better visualization of  $\Delta \ln \mathcal{Z}$ across different datasets, a graphical representation is provided in Fig.~\ref{fig:Bayesian}.

To elucidate the impact of recent DESI-DR2 and SDSS data, we analyze the variation in the theoretical distance prediction of our model and $\Lambda$CDM relative to the observed cosmic distance, as illustrated in Fig. \ref{fig:Distance}. The upper panel displays three different rescaled distance measures for both our model and $\Lambda$CDM, derived using the combined constraints from the OHD+DESI-DR2+Pan+ dataset. Meanwhile, the lower panel presents the normalized residuals between the observational datasets (DESI-DR2 and SDSS)  and the theoretical predictions $-$ our model (\textit{represented by circles}) and $\Lambda$CDM (\textit{represented by crosses}) $-$ expressed in units of the observational uncertainty $\sigma$. Furthermore, to provide a more detailed assessment of each model's agreement with individual BAO measurements, Table~\ref{table:cosmic_distance} summarizes the corresponding comparisons. Overall, our model tends to yield residuals that are comparable to, or smaller than, those of the $\Lambda$CDM model across most redshift bins. Notably, the maximal level of disagreement between DESI-DR2 data and our model remains modest, with a deviation of approximately $-1.867\sigma$ for the measurement of $D_V/(r_d \sqrt{z})$ at $z=0.706$. Similarly, for the SDSS dataset, the largest deviation is approximately $+1.538\sigma$ for the measurement of $D_M/(r_d \sqrt{z})$ at $z=0.700$. Crucially, in both of these instances, the $\Lambda$CDM model exhibits greater tension with the data than our model, reinforcing the consistency of our model across multiple estimators and redshifts. This consistency suggests that our model enhances the agreement with DESI-DR2 data, potentially mitigating small-scale discrepancies and complementing global likelihood analyses. Although the improvements are modest when considered individually, collectively they contribute to a notable enhancement in the overall goodness-of-fit, as shown in Table~\ref{table:cosmic_distance}, where our model achieves a better fit to the OHD+DESI-DR2+Pan+ dataset compared to the concordance $\Lambda$CDM model, with a $\Delta \chi^2 = -7.27$. This result indicates that the preference for our model over standard model stems from its improved capability to reproduce the detailed features observed in the new DESI-DR2 measurements.

\begin{table*}[tpb!]
\begin{center}

\resizebox{2.1\columnwidth}{!}{

\renewcommand{\arraystretch}{1.4}
\tiny
\begin{tabular}{l@{\hspace{0.1cm}} c@{\hspace{0.4cm}} c@{\hspace{0.4cm}} c@{\hspace{0.4cm}}c@{\hspace{0.4cm}}c}
\hline
 \textbf{Distance} & \textbf{Redshift} & \textbf{DESI-DR2} & \textbf{SDSS} & {\bf \texttt{ \bf our model} (\#$\sigma$)} & {\bf$ \Lambda$CDM (\#$\sigma$)}\\ 
 \hline

~\multirow{9}{*}{$D_V(z)/(r_d \sqrt{z})$} & $0.150$ & $-$ & $11.540 \pm 0.44$ & $(+1.147\sigma)$ & $(+1.406\sigma)$ \\
& $0.295$ & $14.622 \pm 0.138$ & $-$ & $-0.275\sigma$ & $+0.357\sigma$ \\
& $0.510$ & $17.812 \pm 0.139$ & $-$ & $+0.507\sigma$ & $+0.694\sigma$ \\
& $0.706$ & $19.102 \pm 0.131$ & $-$ & $-1.867\sigma$ & $-2.008\sigma$ \\
& $0.850$  & $-$ & $19.880 \pm 0.640$ & $(-0.326\sigma)$ & $(-0.388\sigma)$\\
& $0.934$ & $20.406 \pm 0.094$ & $-$ & $+0.022\sigma$ & $-0.491\sigma$ \\
& $1.321$ & $21.101 \pm 0.151$ & $-$ & $-0.013\sigma$ & $-0.382\sigma$ \\
& $1.484$ & $21.388 \pm 0.327$ & $-$ & $+0.697\sigma$ & $+0.548\sigma$ \\
& $2.330$ & $20.484 \pm 0.168$ & $-$ & $-0.203\sigma$ & $-0.075\sigma$ \\

\hline

~\multirow{10}{*}{$D_M(z)/(r_d \sqrt{z})$} & $0.380$ & $-$ & $16.590 \pm 0.280$ & $(-0.455\sigma)$ & $(-0.065\sigma)$ \\
& $0.510$ & $19.027 \pm 0.234$ & $18.710 \pm 0.290$ &   $+1.500\sigma~(+0.117\sigma)$ & $+1.849\sigma~(+0.399\sigma)$\\
& $0.700$ & $-$ & $21.350 \pm 0.390$ &  $(+1.538\sigma)$  &  $(+1.646\sigma)$ \\
& $0.706$ & $20.650 \pm 0.211$ & $-$ &   $-0.732\sigma$    & $-0.537\sigma$\\
& $0.934$ & $22.325 \pm 0.157$ & $-$   &   $-0.986\sigma$  &  $-0.963\sigma$ \\
& $1.321$ & $24.014 \pm 0.277$ & $-$   &  $-0.530\sigma$   &  $-0.645\sigma$  \\
& $1.480$  &  $-$  & $25.230 \pm 0.660$ & $(+0.982\sigma)$ & $(+0.925\sigma)$ \\
& $1.484$  &  $25.047 \pm 0.624$  & $-$  &  $+0.731\sigma$    &    $+0.670\sigma$ \\
& $2.330$ & $25.542 \pm 0.348$ & $24.630 \pm 1.240$ &  $+0.158\sigma~(-0.691\sigma)$  & $+0.077\sigma~(-0.714\sigma)$  \\
& $2.330$ & $-$ & $24.440 \pm 1.110$  &   $(-0.943\sigma)$  &   $(-0.969\sigma)$ \\

\hline

~\multirow{10}{*}{$z D_H(z) / (r_d\sqrt{z})$} & $0.380$ & $-$ & $15.410 \pm 0.470$ & $(+0.984\sigma)$ & $(+0.941\sigma)$ \\ 
& $0.510$ & $15.613 \pm 0.304$ & $15.950 \pm 0.410$  &  $-1.307\sigma~(-0.147\sigma)$ & $-1.538\sigma~(-0.318\sigma)$ \\
& $0.700$ & $-$  & $16.170 \pm 0.440$ &   $ (-1.249\sigma) $  & $(-1.507\sigma)$ \\
& $0.706$ & $16.347 \pm 0.277$ & $-$   &   $-1.384\sigma$  &  $-1.797\sigma$ \\
& $0.934$ & $17.049 \pm 0.187$ & $-$   &   $+1.287\sigma$   &  $+0.616\sigma$  \\
& $1.321$ & $16.293 \pm 0.254$ & $-$   &   $+0.759\sigma$   &   $+0.423\sigma$      \\
& $1.480$ & $-$ & $16.130 \pm 0.670$  & $(+0.673\sigma)$  &  $(+0.583\sigma)$ \\
& $1.484$ & $15.614 \pm 0.629$ & $-$  &    $-0.086\sigma$  &  $-0.181\sigma$     \\
& $2.330$ & $ 13.176 \pm 0.154$ & $13.630 \pm 0.430$ &    $-0.789\sigma~(+0.773\sigma)$  & $-0.328\sigma~(+0.938\sigma)$  \\
& $2.330$ & $ -$ & $13.860 \pm 0.520$  &    $(+1.082\sigma)$   &   $(+1.218\sigma)$ \\

\hline \hline
\end{tabular}
}
\end{center}
\caption{The three rescaled cosmic distance measurement by DESI-DR2 and SDSS, including their associated $\pm 1 \sigma$ errors, are presented (\textit{for SDSS, the results are shown in parentheses}). The agreement between the constraints from OHD+DESI-DR2+Pan+ in our model and the $\Lambda$CDM model with the observed data is assessed, with discrepancies quantified in units of observational uncertainties (\#$\sigma$).}
\label{table:cosmic_distance}
\end{table*}

\section{Summary and Conclusions}
\label{sec-summary}

Revealing the expansion history of the Universe is a major theme in cosmology.  Various cosmological models have been devised; however, according to the existing literature, a consistent cosmological theory in agreement with all the astronomical surveys is yet to be discovered.  Among the list of popular cosmological models, $\Lambda$CDM undoubtedly performs in an  excellent manner with a series of astronomical datasets, but it also has several limitations. Therefore, new cosmological models are equally welcome, provided they fit well with the recent astronomical datasets and are consistent with the 
existing theoretical framework. 

Considering this issue, in the present article we propose a novel parametrized form of the Hubble function (instead of parametrizing the dark energy sector): $H^2/H_0^2 = \Omega_0 (1+z)^{3+\delta} + (1+z)^{\alpha}$, where $\delta$ and $\alpha$ are two constant parameters and $\Omega_0$ may not necessarily correspond to the density parameter of the standard matter sector. At a first glance, one may assume that the above parametrization comes from two noninteracting fluids, one is a noncold DM and one is a DE with constant EoS, but this is not the case, because this resulting Hubble expansion may arise from a series of distinct cosmological models, e.g. noninteracting DE, interacting DE, MG, matter creation, etc.   
Thus, this parametrization  is a novel one having the potentiality to recover many cosmological theories. 
One can quickly identify that for $\delta =0 = \alpha$, the above Hubble expansion law has identical evolution with the $\Lambda$CDM cosmology, but the participating fluids may not be just the non-interacting cold matter sector and the cosmological constant. In a similar way, for $\delta =0$, it represents the $w$CDM-like evolution but again the fluids under action may not correspond to a noninteracting scenario comprising with just a cold matter and a DE with constant EoS.  

In order to perform the observational tests, we have considered several latest astronomical datasets, such as OHD, SNIa (Pan+, DESY5, Union3), BAO (SDSS, DESI-DR1, DESI-DR2), and examined the deviation from the $\Lambda$CDM-like evolution. According to the constraints, we observe a mild deviation from the $\Lambda$CDM-like evolution at slightly more than $1\sigma$, and this evidence increases only for OHD+DESY5+DESI-DR1 and OHD+DESY5+DESI-DR2. This is because of the presence of DESY5 dataset in the combined dataset, and according to recent DESI DR2 results~\cite{DESI:2025zgx}, among all the SNIa datasets, the evidence of dynamical DE mainly increases for DESY5, as highlighted in other studies~\cite{Efstathiou:2024xcq, Huang:2025som, Scherer:2025esj}. Additionally, assuming GR in the background, we also examined the viability of the model in the linear perturbation regime through the growth rate of matter perturbations and we found that the model fits well with the recent compilation of the $f\sigma_8$ measurements.

Further, we assessed the compatibility of the proposed model with the laws of thermodynamics. We found that our model is consistent with the laws of thermodynamics. This strengthens the viability of the proposed parametrization.  Finally, we conducted two model comparison evaluations, namely, information criteria (AIC, BIC, DIC) and Bayesian evidence. We found that AIC and DIC prefer our model over the standard $\Lambda$CDM model across all the datasets, while BIC and Bayesian evidence present mixed behavior, that means, for some datasets, our model is preferred over $\Lambda$CDM and for some other datasets, the reverse situation occurs.  

In summary, as the present model has many interesting features and is flexible to allow a wide variety of cosmological scenarios, therefore, such a parametrized form of the Hubble function can attract the attention of the scientific community to parametrize the dark sector of the Universe without directly parametrizing the individual dark sectors.

\section*{Acknowledgments}
The authors thank the referee for some important comments that resulted in an improved version of the manuscript. 
SP thanks F. B. M. dos Santos for stimulating discussions and for making several comments on the manuscript. 
SP acknowledges the financial support from the Department of Science and Technology (DST), Government of India under the Scheme  ``Fund for Improvement of S\&T Infrastructure (FIST)'' (File No. SR/FST/MS-I/2019/41). We acknowledge computational cluster resources of ICT, Presidency University, Kolkata.

\bibliography{references}

\end{document}